%% file: article_regularized_FP_YC_V4.tex
\documentclass[10pt, twocolumn, twoside]{IEEEtran}

\usepackage{amsmath}
\usepackage{epsfig,amssymb,amsfonts}
\usepackage{amsbsy,bbm}
\usepackage{stmaryrd}
\usepackage[latin1]{inputenc}
\usepackage{verbatim}
\usepackage{graphicx}
\usepackage{graphics}
\usepackage{theorem}
\usepackage{tabularx}
\usepackage{array}
\usepackage{multicol,float}
\usepackage{hhline}
 \usepackage{epstopdf}
\usepackage[T1]{fontenc}
\usepackage{tikz}
\usepackage{tkz-berge}
\usepackage{pgfplots}
\usetikzlibrary{plotmarks}
\usepackage{subfigure}
\usepackage{cite}

\include{definitions}

\def\tred#1{{\color{red}#1}}

\begin{document}
\title{Generalized robust shrinkage estimator and its application to STAP detection problem}

\author{Frédéric Pascal,~\IEEEmembership{Senior Member,~IEEE,}
Yacine Chitour and Yihui Quek%
\thanks{F. Pascal is with SONDRA, Supelec, Plateau du Moulon, 3 rue Joliot-Curie, F-91190 Gif-sur-Yvette, France (e-mail: frederic.pascal@supelec.fr).}%
\thanks{Y. Chitour is with the Laboratoire des Signaux et Syst\`emes, Supelec, Plateau du Moulon, 3 rue Joliot Curie, F-91192 Gif-sur-Yvette Cedex, France (e-mail: yacine.chitour@lss.supelec.fr).}
\thanks{Y. Quek is a B.Sc student in Physics and Applied Mathematics at the Massachusetts Institute of Technology, 77 Massachusetts Ave, Cambridge, USA}
\thanks{This work has been partially supported by the DGA grant N° 2013.60.0011.00.470.75.01 for F. Pascal and by the ICODE institute for F. Pascal and Y. Chitour.}}

\markboth{Submitted to IEEE Trans. on Signal Processing}{Pascal \MakeLowercase{\textit{et al.}}: Covariance matrix estimation: generalized robust shrinkage estimator versus Tyler estimator}

\maketitle

\begin{abstract}
Recently, in the context of covariance matrix estimation, in order to improve as well as to regularize the performance of the Tyler's estimator \cite{tyler1987distribution} also called the Fixed-Point Estimator (FPE) \cite{pascal2008covariance}, a "shrinkage" fixed-point estimator has been {originally introduced in \cite{abramovich2007diagonally}}. First, this work extends the results of \cite{chen2011robust,wiesel2012unified} by giving the general solution of the "shrinkage" fixed-point algorithm. Secondly, by analyzing this solution, called the generalized robust shrinkage estimator, we prove that this solution converges to a unique solution when the shrinkage parameter $\beta$ (losing factor) tends to 0. This solution is exactly the FPE with the trace of its inverse equal to the dimension of the problem. This general result allows one to give another interpretation of the FPE and more generally, on the Maximum Likelihood approach for covariance matrix estimation when constraints are added. Then, some simulations illustrate our theoretical results as well as the way to choose an optimal shrinkage factor. Finally, this work is applied to a Space-Time Adaptive Processing (STAP) detection problem on real STAP data.
\end{abstract}


\section{Introduction}
\label{sec:intro}

In the statistical signal processing area, the problem of covariance matrix estimation is an active topic of research \cite{wiesel2012unified, pascal2012mggd, abramovich2012regularized, mahot2013asymptotic, ollila2012complex, ollila2012compound, chen2011robust, chitour2008exact, pascal2008covariance}. From an application point of view, a better accuracy in terms of covariance matrix estimation directly involves an improvement of the system performance in terms of estimation, detection and/or classification, as shown in radar applications or in Direction-Of-Arrival estimation problems (see e.g. \cite{mahot2013asymptotic, ollila2012complex} and references therein). Until recent years, the classical assumption used for data modelling was the Gaussian model that provides the well-known Sample Covariance Matrix (SCM) estimator as the Maximum Likelihood (ML) estimator for the data covariance matrix. However, in many practical cases, the SCM suffers from major drawbacks, as for instance in adaptive radar and sonar processing \cite{Kay98} since its performance can be strongly degraded. This is also the case in the presence of non-Gaussian, impulsive and/or heterogeneous noise as well as in the presence of outliers (see e.g. \cite{Mahot10} and references therein). To fill these gaps, a general framework on robust estimation theory has been extensively studied in the statistical community in the 1970s following the seminal works of Huber, Hampel and Maronna \cite{huber1964robust, Huber72, Hamp74, Maronna76}. The multivariate real case introduced by Maronna in \cite{Maronna76} has been recently extended by Ollila to the complex case \cite{ollila2012complex, mahot2013asymptotic, pascal2008covariance} more adapted for signal processing applications.\\

Under this robust theory framework, most of recent works in covariance matrix estimation considers the broader class of Complex Elliptically Symmetric (CES) distributions, originally introduced by Kelker \cite{kelker1970distribution}, which encompasses the Spherically Invariant Random Vectors (SIRV) \cite{Yao73} as well as the Multivariate Generalized Gaussian Distributions (MGGD) \cite{Gomez1998}. \cite{ollila2012complex} provides a  complete review on CES applied to array processing. We will refer to this paper for main results on CES. From a signal processing point of view, an important contribution of this class of distributions is that the covariance matrix does not necessary exist, which is the case for instance in the multivariate Cauchy distribution whose variance is infinite. Thus, one can always consider the so-called scatter matrix that is always well defined, contrary to the covariance matrix. In the case of finite second-order moment, the covariance matrix is equal to the scatter matrix, up to a scale factor (see \cite{ollila2012complex} for more details). One important consequence is that one can always estimate the scatter matrix instead of the covariance matrix. Moreover, for applications that are invariant to a scale factor, like for instance DOA estimation with the MUltiple SIgnal Classification (MUSIC) \cite{schmidt1979multiple} or detection using the Adaptive Normalized Matched Filter (ANMF), 
{firstly introduced by \cite{Conte95} and analyzed in \cite{liu2011acfar, kraut1999cfar, Kraut01, Pascal04-1}}, the resulting performance is the same.\\

In this context of covariance matrix estimation, to improve the estimator performance as well as to deal with under-sampling cases (i.e. when the number of sample is less than the dimension of the data), a common regularization approach has been widely studied, the diagonally loaded approach originally introduced by \cite{abramovich1981controlled, carlson1988covariance} and applied to the SCM. More recently, a shrinkage has been proposed in \cite{ledoit2004well} but also applied to the SCM. To deal with non-Gaussian models and to have a robust approach, this shrinkage has been applied to the Tyler's estimator \cite{tyler1987distribution} to obtain a "shrinkage" fixed-point estimator (FPE), {originally introduced by \cite{abramovich2007diagonally}} for the case where the number of samples is less than the dimension. {A rigorous proof of existence, uniqueness and convergence of the associated recursive algorithm has been given by \cite{chen2011robust} in the case where a trace normalization has been added. Moreover, in \cite{wiesel2012unified}, a similar shrinkage fixed-point estimator has been analyzed by including a penalized term on the trace of the inverse.} Then, in \cite{abramovich2013covariance, abramovich2013regularized1, besson2013regularized2}, this estimator has been used with the Expected Likelihood approach. However, in all these work, no general proof for the existence and the uniqueness of this "shrinkage" FPE is provided. Only the particular case where the trace of the estimator is fixed is analyzed. \\ 

To fill this gap, this work provides the general solution of this FP problem, even in the case of under sampling, i.e. when the number $N$ of samples is less than the dimension $m$ of the observations. Moreover, this "shrinkage" FPE is compared to the classical FPE and interestingly, it provides a simple way to built a unique FPE of the covariance matrix. Finally, the proof relies on the analysis of a continuous function that can be seen as a Likelihood Function (LF) that generalizes the LF of the so-called Angular-Complex Gaussian (ACG) distributions.\\

The second part of this paper is devoted to the analysis of the "shrinkage" FPE in a Space-Time Adaptive Processing (STAP) context \cite{Ward94, Klemm02}. For that purpose, the shrinkage parameter is studied in order to provide the better detection performance. Regarding the work of \cite{abramovich2013regularized1}, this paper considers the so-called over-sampled case, which means that the number $N$ of samples is greater than the dimension $m$ of the observations. However, some preliminary results on the under-sampled case will be provided when applying proposed approach on STAP data. These results are linked with those of \cite{besson2013regularized2} that also considers the under-sampled case, i.e. where the number $N$ of samples is less than the dimension $m$ of the observations.\\

The paper is organized as follows: section \ref{sec:back} presents the estimation context while section \ref{sec:main} contains the main contribution of this work, i.e. the derivation of the Shrinkage FPE in a general context. First part of section \ref{sec:simus} is devoted to the analysis of the Shrinkage FPE through Monte Carlo simulations, and then this estimate is applied to a STAP detection problem on a real set of data. Section \ref{sec:conclu} draws the conclusions and gives some outlines for further work. For the clarity of the presentation, some parts of the proofs are postponed in the Appendix section.\\

The following convention is adopted: italic indicates a scalar quantity, lower case boldface indicates a vector quantity and upper case boldface a matrix. $^T$ denotes the transpose operator and $^H$ the transpose conjugate. $E[.]$ is the expected value operator and $\text{Tr(.)}$ denotes the trace operator. $\CN (\ag, \Mg)$ is a complex Gaussian distribution with a mean vector $\ag$ and a covariance matrix $\Mg$. $\Ig$ is the identity matrix with appropriate dimension.

\section{Background}\label{sec:stat}
\label{sec:back}
In the context of covariance matrix estimation, this paper focuses on two particular estimators: the FPE or Tyler's estimator and the shrinkage fixed point estimator also called the diagonally loaded fixed point estimator. Let us consider a $N$-sample $(\xg_1, \hdots, \xg_N)$ of independent and identically distributed (i.i.d.) $m$-variate random vectors with covariance matrix $\Sig_0$ if it exists, else $\Sig_0$ is the scatter matrix.  The FPE or Tyler's estimator \cite{tyler1987distribution, pascal2008covariance} which is defined as the solution of the following fixed-point equation:
\begin{equation}
\label{FPE}
\Sig = f(\Sig),
\end{equation}
where the map $f$ is defined over the positive definite hermitian matrices of size $m$ by
\begin{equation}\label{map}
 f(\Sig)= \cfrac m N \disp \sum_{n=1}^N\cfrac{\xg_n \xg_n^H}{\xg_n^H \Sig^{-1} \xg_n}.
 \end{equation}
As shown in \cite{pascal2008covariance} for the complex case, the solution exists and is unique up to a scale factor. Moreover, the associated recursive algorithm defined by
\begin{equation}
\label{FPE-algo}
\left\{
\begin{array}{lll}
\widetilde{\Sig}_{(k+1)} & = & \cfrac m N \disp \sum_{n=1}^N\cfrac{\xg_n \xg_n^H}{\xg_n^H \widehat{\Sig}_{(k)}^{-1} \xg_n}\\
\widehat{\Sig}_{(k+1)} & = & \cfrac{m}{\Tr\be\widetilde{\Sig}_{(k+1)}\en} \, \widetilde{\Sig}_{(k+1)}
\end{array}
\right.
\end{equation}
converges towards the solution which respects the constraints that its trace is equal to $m$, whatever the initialization matrix $\Sig_{(0)}$.\\

It is important to notice that the constraint on the trace, used here for identifiability considerations, is only considered in the recursive algorithm to obtain a unique solution. {However, this solution is not the general solution of Eq. \eqref{FPE}, since this general solution is not unique but is defined up to a scale constant as shown in \cite{pascal2008covariance}. More precisely, under the constraint $\Tr(\Sig)=m$, it is proved in \cite{pascal2008covariance} that a particular solution exists, is unique and can be achieved by using the recursive algorithm defined in Eq. \eqref{FPE-algo}. In the following, this point will be more detailed concerning the shrinkage fixed point estimator, since it is proved in this work that no additional constraint is required to obtain the uniqueness of a solution of the shrinkage fixed point equation.}\\

Let us consider now the shrinkage (diagonally loaded) fixed point originally introduced in {\cite{abramovich2007diagonally}, fully analyzed in \cite{chen2011robust}} and defined as the solution of the following fixed point equation, for $\beta \in [0,1]$
\begin{equation}
\label{FPE-shrinkage}
\Sig(\beta) = (1-\beta)\, \cfrac m N \disp \sum_{n=1}^N\cfrac{\xg_n \xg_n^H}{\xg_n^H \Sig(\beta)^{-1} \xg_n} + \beta\, \Ig.
\end{equation}
Notice that no proof of existence and uniqueness of a solution of Eq. \eqref{FPE-shrinkage} is given in \cite{chen2011robust} as stated by Theorem 1 of \cite{chen2011robust}. Actually, it is proved that the following recursive algorithm converges to a unique solution whatever the initialization:
\begin{equation}
\label{FPE-shrinkage-algo}
\left\{
\begin{array}{lll}
\widetilde{\Sig}_{(k+1)} & = & (1-\beta)\, \cfrac m N \disp \sum_{n=1}^N\cfrac{\xg_n \xg_n^H}{\xg_n^H \widehat{\Sig}_{(k)}^{-1} \xg_n} + \beta\, \Ig\\
\widehat{\Sig}_{(k+1)} & = & \cfrac{m}{\Tr\be\widetilde{\Sig}_{(k+1)}\en} \, \widetilde{\Sig}_{(k+1)}
\end{array}
\right.
\end{equation}

{In \cite{wiesel2012unified}, a similar shrinkage fixed point has been proposed and is defined by
\begin{equation}
\label{FPE-shrinkage-AW}
\tilde{\Sig}(\beta) = (1-\beta)\, \cfrac m N \disp \sum_{n=1}^N\cfrac{\xg_n \xg_n^H}{\xg_n^H \tilde{\Sig}(\beta)^{-1} \xg_n} + \beta\, \cfrac{m}{\Tr(\tilde{\Sig}(\beta)^{-1})}\Ig.
\end{equation}
The aim of this paper is to analyze the fixed-point scheme defined by Eq.\eqref{FPE-shrinkage}. This is the purpose of the next section.}

\section{Main contribution}
\label{sec:main}
{One of the main contributions of this paper is to prove that Eq. \eqref{FPE-shrinkage} admits a unique solution for $\beta\in (\bar{\beta},1]$, where $\bar{\beta}:=\max(0, 1-N/m)$ and that this solution can be achieved by a similar algorithm to \eqref{FPE-shrinkage-algo} but without the step that imposes the trace of the solution equal to $m$. Before turning into the proof of such a result, let us notice that, interestingly, we have the next proposition.
\begin{Prop}
\label{prop-trace}
If Eq.~\eqref{FPE-shrinkage} admits a solution $\Sig$ for some $\beta \in (0,1]$, thus $\Sig$ verifies the following constraint:
\begin{equation}\label{trace-inv-constaint}
\Tr\be \Sig^{-1}\en = m.
\end{equation}
\end{Prop}
\begin{IEEEproof}
Let us whiten Eq. \eqref{FPE-shrinkage} by $\Sig^{-1}$:
\begin{equation*}
\Ig = (1-\beta)\, \cfrac m N \disp \sum_{n=1}^N\cfrac{\Sig^{-1/2}\xg_n \xg_n^H \Sig^{-1/2}}{\xg_n^H \Sig^{-1} \xg_n} + \beta\, \Sig^{-1}.
\end{equation*}
By setting $\zg_n = \Sig^{-1/2} \xg_n/\sqrt{\xg_n^H \Sig^{-1} \xg_n}$, which is of unit norm,
one obtains
\begin{equation}\label{eq00}
\Ig = (1-\beta)\, \cfrac m N \disp \sum_{n=1}^N \zg_n \zg_n^H + \beta\, \Sig^{-1},
\end{equation}
and by taking the trace, one has
\begin{equation*}
m = (1-\beta)\, \cfrac m N \disp \sum_{n=1}^N \Tr\be\zg_n \zg_n^H\en + \beta\, \Tr\be\Sig^{-1}\en.
\end{equation*}
Since each $\zg_n$ is of unit norm, one has $\Tr\be\zg_n \zg_n^H\en = 1$ and the above equation reduces to
\begin{equation*}
m = (1-\beta)\, m + \beta\, \Tr\be\Sig^{-1}\en,
\end{equation*}
which concludes the proof for any $\beta$ different from 0.\\
\end{IEEEproof}
Consequently, there is no reason for a solution to simultaneously verify $\Tr\be \Sig^{-1}\en = m$ and $\Tr\be \Sig\en = m$ as it is provided by the algorithm of \cite{chen2011robust}, recalled in Eq.~\eqref{FPE-shrinkage-algo}. Moreover, the way proposed by \cite{wiesel2012unified} that penalizes on the trace of the inverse is not necessary since this constraint is naturally satisfied.\\\\
The second contribution of this paper consists in showing, when $m/N<1$, that the map $\beta\mapsto \Sig(\beta)$ defined on $(0,1]$ admits a limit as $\beta$ tends to zero and thus converges to the unique fixed point of Eq. \eqref{FPE} whose inverse has its trace equal to $m$.
}
The following theorem proves the existence and the uniqueness of a solution of the fixed point Eq. \eqref{FPE-shrinkage} for any $\beta$ between 0 and 1 (bounds are included) when $N>m$, but also in the case where $N\leq m$ for particular values of $\beta$. On the other hand, a simpler recursive algorithm is provided, whose convergence to the solution is ensured.

We use $\D$ to denote the set of Positive Definite Symmetric (PDS) matrices of size $m$. 
\begin{Thm}\label{thm-exist-shrinkage}
The following fixed point equation 
\begin{equation}\label{fpe-1}
\Sig = (1-\beta)\, \cfrac m N \disp \sum_{n=1}^N\cfrac{\xg_n \xg_n^H}{\xg_n^H \Sig^{-1} \xg_n} + \beta\, \Ig,\quad (\Sig,\beta)\in\D\times (0,1],
\end{equation}
admits a solution if and only if $\beta\in (\bar{\beta},1]$, where $\bar{\beta}:=\max(0, 1-N/m)$, and in that case, the solution is unique and denoted $\Sig(\beta)$. 

Moreover, when $m/N<1$, $\lim_{\beta\rightarrow 0}\Sig(\beta)$ exists and is equal to the unique fixed point of Eq.\eqref{FPE} whose inverse has a trace equal to $m$.
\end{Thm}

Let us notice that Theorem \ref{thm-exist-shrinkage} provides the main result of this work. It directly implies that the unique solution of Eq. \eqref{FPE-shrinkage} tends to a particular Tyler's estimator as $\beta$ tends to 0, i.e. the unique estimator that verifies that the trace of its inverse is equal to $m$.\\

In the course of the argument of Theorem \ref{thm-exist-shrinkage}, we will  be considering  the functions $F:\D \times [0,1]\rightarrow (\RR^+)^{\ast}$ and $f:\D \times [0,1]\rightarrow\D$
given by 
\begin{equation}
\label{F-shrinkage}
F(\Sig,\beta):=\cfrac{\exp(-N\beta \Tr(\Sig^{-1})) }{\det(\Sig)^N} \disp  \prod_{n=1}^N  (\xg_n^{\top}  \Sig^{-1} \xg_n)^{-m(1-\beta)}
\end{equation}
\begin{equation}
\label{f-shrinkage}
f(\Sig,\beta):= (1-\beta) \, \cfrac{m}{N}\,\disp \sum_{n=1}^N \cfrac{\xg_n \, \xg_n^{\top}}{\xg_n^{\top}  \Sig^{-1} \, \xg_n} + \beta\, \Ig.
\end{equation}
The extension to the complex numbers follows straightforwardly, as proved in \cite{pascal2008covariance}. In the above functions, {$\beta$ appears as an argument but it will be fixed ($\beta \in (\bar{\beta},1]$) in the proof.} These notations will allow one to maximize the function $F(\cdot,\beta)$, that can be assimilated as a Likelihood function, with respect to (w.r.t.) $\Sig$. As mentioned in \cite{abramovich2013regularized1,besson2013regularized2}, an important challenge when dealing with the shrinkage estimator is to find a "good" shrinkage parameter. Discussions on the theorem results are provided after the proof in remark \ref{rem-main-thm}. \\

\begin{IEEEproof} 
The proof is divided into two parts according to the value of the ratio $m/N$.
\subsection{Case where $m/N<1$}
The proof strategy is similar to that of \cite{chitour2008exact}. More precisely, we will first prove that, for every $\beta\in (0,1]$, Eq. \eqref{FPE-shrinkage} has solutions by combining two facts: $(a)$ solutions of Eq. \eqref{FPE-shrinkage} are exactly the critical points of $F(\cdot,\beta)$ and $(b)$ $F(\cdot,\beta)$ admits a unique strict global maximum on $\D$. In a second step, we will show that every critical point $F(\cdot,\beta)$ in $\D$ must be a strict local maximum and we hence conclude the first part of Theorem \ref{thm-exist-shrinkage} relying on a 
 topological argument.
As for the second one, this will result the study of the map $\beta\mapsto \Sig(\beta)$ as $\beta$ tends to zero.

In the sequel, we use $F_\beta$ and $f_\beta$ to denote respectively the maps over $\D$ given by $F(\cdot,\beta)$ and $f(\cdot,\beta)$.
Note that the maps $F_0$ and $f_0$ have been studied in detail in \cite{pascal2008covariance}.
Let us now consider the "$\log$-likelihood function"
\begin{multline*}
\log(F_\beta)(\Sig) = -N\log(\det(\Sig)) -N\beta\Tr(\Sig^{-1}) \\ - m(1-\beta)\disp\sum_{n=1}^N \log(\xg_n^{\top}  \Sig^{-1} \xg_n). 
\end{multline*}
By differentiation w.r.t $\Sig$, one obtains
\begin{equation}
\nabla \log(F_\beta)(\Sig)= -N\Sig^{-1}\big(\Sig-f_\beta(\Sig)\big)\Sig^{-1},
\end{equation}
where $\nabla \log(F_\beta)(\Sig)$ is the unique symmetric matrix such that $dF_\beta(\Sig)(Q)=\Tr(\nabla \log(F_\beta)(\Sig)Q)$ for every symmetric matrix $Q$. We therefore trivially conclude that the fixed points of $f(\cdot,\beta)$ are exactly the critical points of $F_\beta$.

We next state the following proposition.
\begin{Prop}
\label{prop-comparaison}
Assume that $m/N<1$. For every $(\Sig,\beta)\in \D\times [0,1]$, one has that
$(a)$ $F_\beta$ can be extended by continuity on the boundary of $\D$ by zero; $(b)$ $F_\beta(\Sig)$ tends to zero as $\Vert \Sig\Vert$ tends to infinity. Hence,  $F_\beta$ admits a global maximum in $\D$.
\end{Prop}
{\begin{IEEEproof} 
The proof has been postponed in Appendix \ref{app-prop-comparaison}.\\
\end{IEEEproof}} 

We next prove that every critical point of $F_\beta$ in $\D$ must be a local strict maximum. We then immediately conclude the argument for the existence of a solution of  Eq. \eqref{FPE-shrinkage}. 

\begin{Prop}
\label{prop-hessien}
For every $\beta\in (0,1]$, if $\Sig\in \D$ is a critical point of $F_\beta$, then 
\begin{equation}\label{est1}
\frac{Hess_\beta(\Sigma)(Q)}{N\beta F_\beta(\Sig)}\leq -\beta\Tr(Q\Sig^{-2}Q\Sig^{-2}),
\end{equation}
where $Hess_\beta(\Sig)$ is the Hessian of $F_\beta$ at $\Sig$. One deduces at once that 
$\Sig$ must be a local strict maximum.
\end{Prop}
{
\begin{IEEEproof}
The proof has been postponed in Appendix \ref{app-prop-hessien}.\\
 \end{IEEEproof}
}
As for the uniqueness of the solution of  Eq. \eqref{FPE-shrinkage}, we argue by contradiction. Assuming that two such solutions $\Sigma_1$ and $\Sigma_2$ exist in $\D$,  we can apply the mountain-pass theorem to the functional $1/F_\beta$  (\cite{struwe2008variational}) since the latter functional tends to infinity as one approaches the boundary of $\D$. We thus obtain the existence of a saddle point of $F_\beta$ in $\D$, which is not possible according to Proposition
\ref{prop-hessien}.

We now turn to the second statement in Theorem \ref{thm-exist-shrinkage}, namely the fact that $\lim_{\beta\rightarrow 0}\Sig(\beta)$ exists and belongs to $\D$. First notice that if $\widehat\Sig$ is an accumulation point of $\Sig(\beta)$ as $\beta$ tends to zero such that $\widehat\Sig\in \D$, then $\widehat\Sig$ must be a fixed point of $f_0$ whose inverse has trace equal to $m$. According to Theorem IV.1 of \cite{pascal2008covariance}, the latter is unique. To prove the thesis, it is therefore sufficient to prove that $\Sig(\beta)$ is lower bounded, for $\beta$ small enough, by a element of $\D$. That last statement would follow from an upper bound for $\det(\Sig(\beta))$, for $\beta$ small enough, since $\Tr(\Sig(\beta)^{-1})=m$ implies that $\Sig(\beta)\geq 1/m\ \Ig $.

This is the object of the following proposition.

\begin{Prop}
\label{prop-LB}
There exists $C_0>0$ such that, for every $\beta\in (0,1)$, $\det(\Sig(\beta))\leq C_0$.
\end{Prop}
{\begin{IEEEproof} 
The proof has been postponed in Appendix \ref{app-prop-LB}.\\
\end{IEEEproof}
}

The proof of Theorem \ref{thm-exist-shrinkage} when $m/N<1$ is then complete. Let us now turn to the case where $m/N\geq 1$.
\subsection{Case where $m/N\geq 1$}
We have the following result.
\begin{Prop}
\label{prop-ext}
The fixed point equation \eqref{fpe-1} admits a unique solution if and only if  $(1-\beta)m/N<1$.
\end{Prop}

{\begin{IEEEproof} 
The proof has been postponed in Appendix \ref{app-prop-ext}.\\
\end{IEEEproof}
}

This concludes the proof of Theorem \ref{thm-exist-shrinkage}.
\end{IEEEproof}

Some comments on the results of Theorem \ref{thm-exist-shrinkage} are given in the following remark.
\begin{rem}\label{rem-main-thm}
Two main points are involved by Theorem \ref{thm-exist-shrinkage}:

\begin{itemize}
\item First, when $m/N>1$, the parameter $\beta$ should be greater than $1-N/m$, which is a very intuitive condition. Indeed, the greater the ratio of m/N, the more \textit{a priori} information is needed or equivalently, the stronger the regularization has to be. This implies for this shrinkage FP algorithm that the weight applied on the identity matrix has to be higher.
{\item Then, the proposed shrinkage FPE is unique, which differs from the one defined by Eq. \eqref{FPE-shrinkage-AW} that is unique only up to a scaling factor. Actually, the consequence of the penalization function used in \cite{wiesel2012unified} is to removed the natural trace constraint.}
\item Moreover, the approach used to prove Theorem \ref{thm-exist-shrinkage} relies on the analysis of a likelihood function (LF), which can be seen as a generalization of the complex angular Gaussian (AG) distributions detailed in \cite{ollila2012complex}, due to the parameter $\beta$. Of course, when $\beta=0$, one retrieves the classical AG distribution. Interestingly, this LF can now be maximized w.r.t the CM but also w.r.t to $\beta$, to obtain a way to optimally set the $\beta$ parameter. Unfortunately, as shown in the Appendix {\ref{app-prop-maxglob}} by Proposition \ref{prop-maxglob}, the Log-likelihood function is convex and so the maximum is obtained for $\beta = 0$ or $\beta=1$. 
{\item Following previous remark, a way to find an optimal (in the sense of minimizing the Mean Square Error) value of $\beta$ has been recently proposed in \cite{couillet2014large} by means of large Random Matrix Theory.}
\end{itemize}
\end{rem}

\section{Simulations}
\label{sec:simus}

\input{fig1.tex}

This section is divided into two parts. First, the algorithm behaviour of the Shrinkage estimate is analysed and compared to the FPE, only for estimation purposes. Moreover, the convergence of the Shrinkage estimate towards the particular FPE that verified $\Tr(\widehat{\Sig}^{-1})=m$, proved by Theorem \ref{thm-exist-shrinkage}, is illustrated when the shrinkage parameter $\beta$ tends to 0. 

Then, the Shrinkage estimate is used in a STAP application and its performance is compared to the one of the FPE, the standard Sample Covariance Matrix (SCM) and the Diagonally Loaded-SCM (DL-SCM).

\subsection{Shrinkage FPE algorithm analysis}

In these simulations, the true covariance matrix $\Sig = \alpha \Gr M$ is defined through a correlation coefficient $\rho$ as follows:
\begin{equation}
\Gr M_{ij} = \rho^{|i-j|}, \text{ for } \rho \in (0,1)
\end{equation}
where $\alpha$ is a scalar that ensures $\Tr(\Sig^{-1}) = m$, i.e. $\alpha = \Tr(\Gr M^{-1})/m$, to be coherent with the theoretical results and the discussions on the trace constraint. Consequently, the particular FPE that is studied is the one whose trace of its inverse is equal to $m$. But, of course, for application purposes with a priori knowledge on this coefficient, the recursive algorithm can be modified without changing the results (convergence to a unique solution) as it is proved in Theorem \ref{thm-exist-shrinkage}.  This definition allows to use covariance matrices close to the identity matrix (for $\rho$ close to 0) as well as bad conditioned covariance matrices when $\rho$ is close to 1. The samples are zero-mean generated from Gaussian distribution with covariance matrix $\Sig$. Let us notice that the result are still valid for any Spherically Invariant Random Vectors (SIRV) since, in the definition of both estimators (Shrinkage estimator and FPE), they do not depend on a scalar factor that multiplies the samples $\xg_n$.

Furthermore, the dimension $m$ of the data is settled to be equal to 12, the number $N$ of samples is equal to 24, 48 or 200 and to assess estimates performances, the Normalized Mean Square Error (NMSE) is used. \\

Figures \ref{err-rho=0.01}, \ref{err-rho=0.5} and \ref{err-rho=0.99} show the NMSE versus the regularized parameter $\beta$ for the Shrinkage estimator as well as for the FPE, for different values of the correlation coefficient, i.e. $\rho = 0.01, 0.5 \text{ and } 0.99$ and for a number of samples $N$ equal to 24, i.e. twice the dimension of the observations. Let us first notice that in all scenarios, the Shrinkage estimator outperforms the FPE, particularly when $\rho$ is closed to 0, since the regularization enforces to provide an estimate close to the identity matrix. But, even for large values of $\rho$, the NMSE of the Shrinkage estimator is lower than the one of the FPE. Another interesting result is that the optimal\footnote{in the sense of the particular criterion used in this work, i.e. the NMSE} regularized parameter changes according to the true covariance matrix: larger is the correlation coefficient $\rho$, smaller is the optimal regularized parameter, and conversely.\\ 

\input{fig2.tex}
\input{fig2-b.tex}

This behavior is also present on Figures \ref{err-rho=0.01-N=48}, \ref{err-rho=0.5-N=48} and \ref{err-rho=0.99-N=48} where the number of samples is equal to $N=48$. But interestingly, for a larger $N$, the Shrinkage estimator does not always outperform the FPE. Moreover, when N increases, the optimal value of $\beta$ will  be equal to $0$, i.e. the Shrinkage estimator will be reduced to the FPE, as illustrated on Figure \ref{fig-2-N=200} for $\rho=0.5$ and 0.99  and $N=200$. This can be explained by the fact that the FPE is a consistent ML  estimator, as shown in \cite{ollila2012complex,pascal2008performance} associated to a particular distribution of the sample while, as shown by equation \eqref{F-shrinkage}, the Shrinkage estimator does not result from any known distribution. Furthermore, this confirms what has been shown previously, which is that maximizing the LF associated to the Shrinkage estimate on both $\Sig$ and $\beta$ is not a good way for optimizing on the regularized parameter $\beta$ since the optimization will provide $\beta = 0$ or $\beta=1$.

Conversely, when the number of samples $N$ is small, the regularization of the estimator plays an important role (as it is designed for it) and allows to significantly improve the performance, again in the sense of the NMSE. Besides, as shown in the following application, the regularization also allows one to deal with problem where $N$ is smaller than the dimension $m$, which is impossible by using the FPE since it requires a matrix inversion in its definition.\\

Finally, Figure \ref{fig-4} illustrates the convergence of the Shrinkage estimator towards the FPE when $\beta$ tends to 0, by plotting the following criterion: $C_1(\beta)=\|\Sig(\beta)-\Sig_{FP}\|_F/\|\Sig_{FP}\|_F$ for different values of $\rho$ and for $m=3$ and $N=12$.

\input{fig3-cv.tex}

\subsection{Application to real STAP data}
This section is devoted to the analysis of the Shrinkage estimator performance on a real STAP data set. This performance are analyzed through the target detection problem. {STAP \cite{Ward94, Klemm02} is a recent technique used in airborne phased array radar to detect moving targets embedded in an interference background such as jamming or strong clutter. While conventional radars are capable of detecting targets both in the time domain related to target range and in the frequency domain related to target velocity, STAP uses an additional domain (space or information collected by an antennas array) related to the target angular localization.  The joint processing of these space-time data, by appropriate two-dimensional adaptive filtering methods, allows stronger interference/clutter rejection and therefore improved target detection.} \\

The STAP data are provided by the French agency DGA/MI\footnote{The authors are grateful to DGA/MI for providing them this set of real data}: the clutter is real but the targets are synthetic. The number of sensors is $S=4$ and the number of coherent pulses can be up to $M=64$. The centre frequency and the bandwidth are respectively equal to $f_0=10~$GHz and $B=5~$MHz. The radar velocity is given by $V=100~$m/s. The inter-element spacing is $d=0.3~$m and the pulse repetition frequency is $f_r=1~$kHz. The Clutter to Noise Ratio (CNR) is equal to 20~dB. The maximum number of samples available for all scenarios is $N=408$. In the different scenarios, targets with a Signal to Clutter Ratio (SCR) of -5~dB are present.\\

\begin{figure}[!htp]
\begin{center}
\includegraphics[width=1\linewidth,height=12cm]{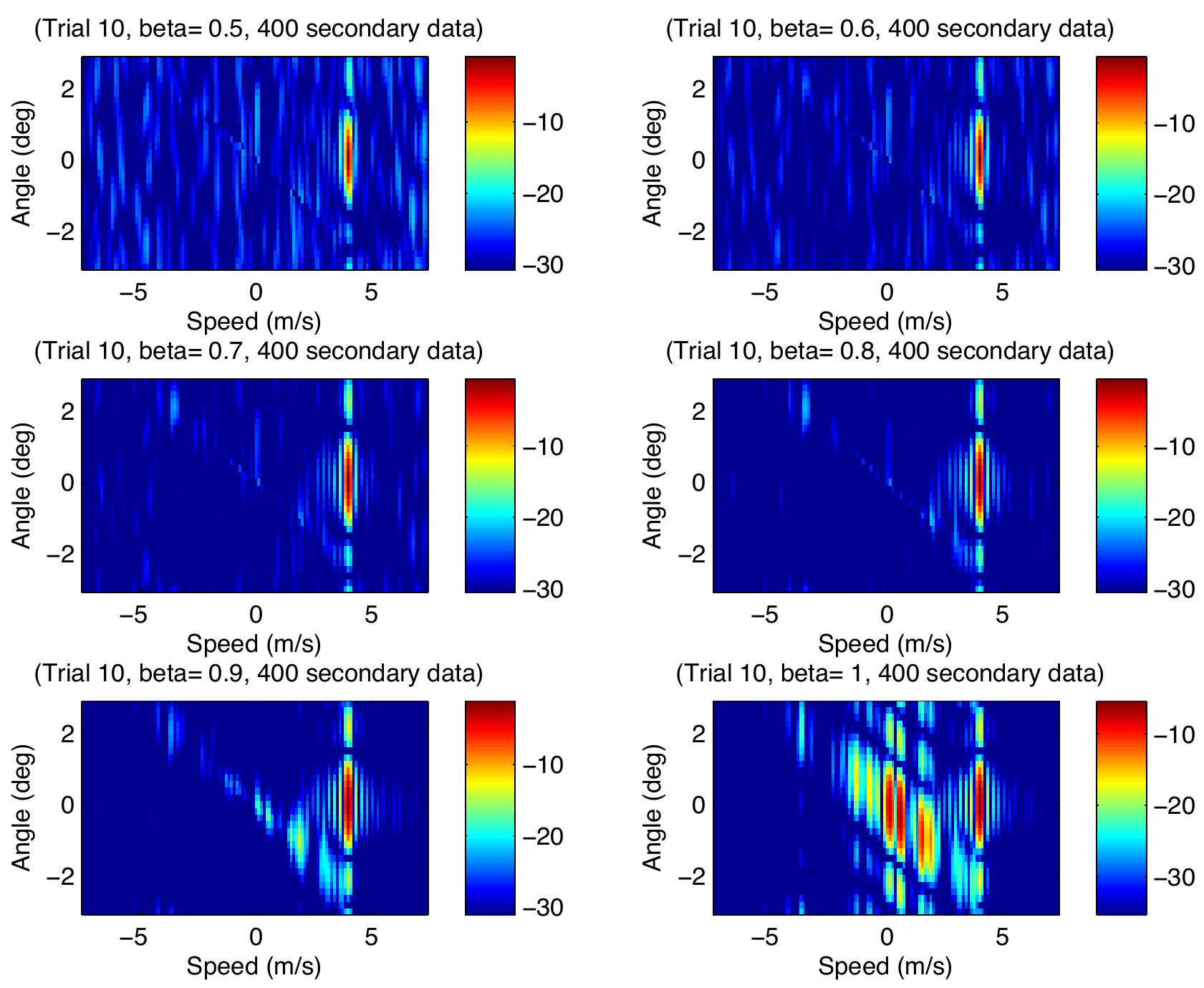}
\caption{$\log_{10}(\Lc_{S-FPE})$ for different values of $\beta$: angle/speed map with parameters $m=256$, $N=400$. The colour scale is in dB with a dynamic of 30 dB from the maximum value of $\log_{10}(\Lc_{S-FPE})$}
\label{fig-STAP-SFPE-N400}
\end{center}
\end{figure} 
\begin{figure}[!htp]
\begin{center}
\includegraphics[width=1\linewidth,height=12cm]{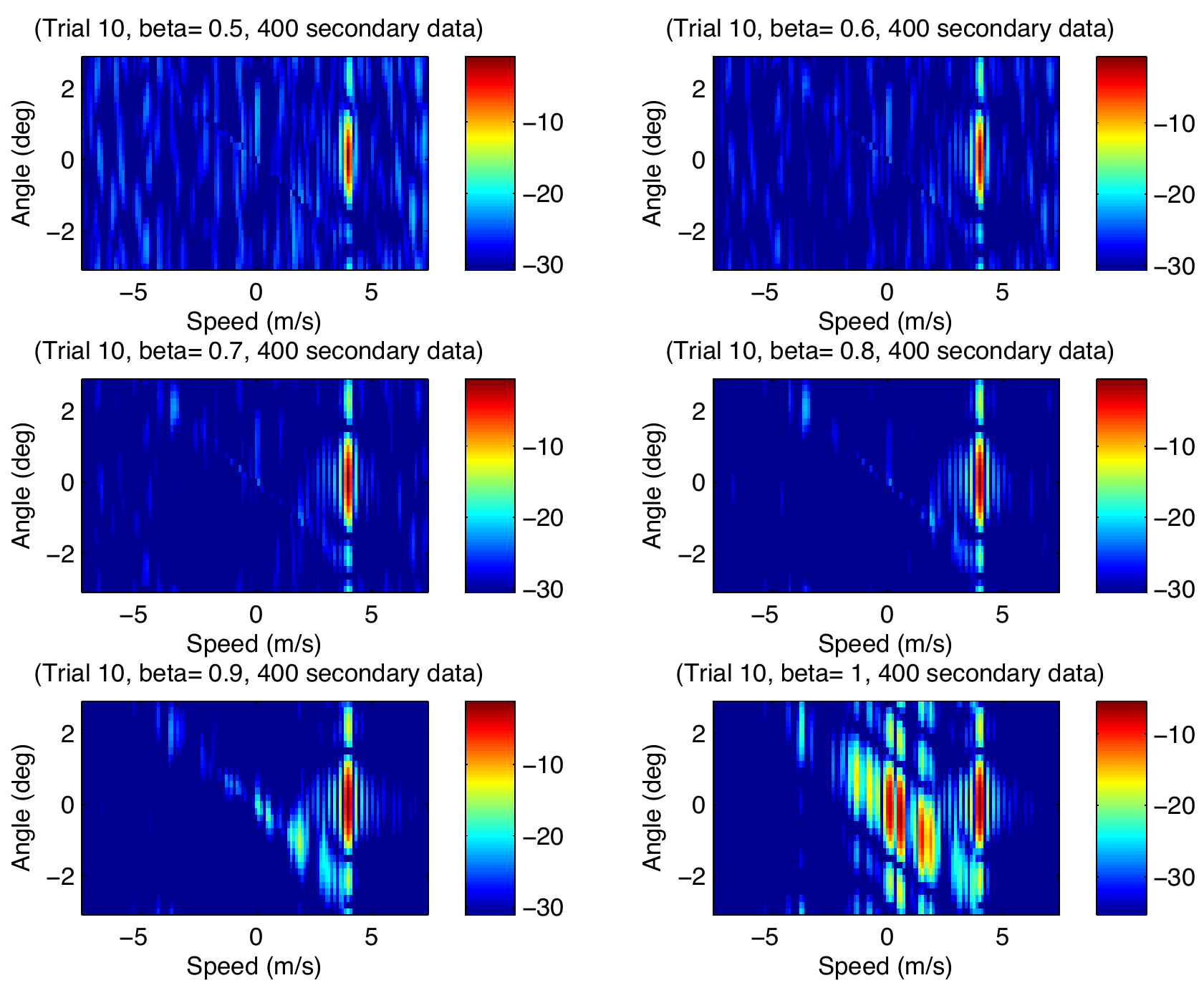}
\caption{$\log_{10}(\Lc_{S-FPE-W})$ for different values of $\beta$: angle/speed map with parameters $m=256$, $N=400$. The colour scale is in dB with a dynamic of 30 dB from the maximum value of $\log_{10}(\Lc_{S-FPE-W})$}
\label{fig-STAP-SFPE-W-N400}
\end{center}
\end{figure} 
The data under study contains 3 synthetic targets ($4~$m/s, $0\deg$, cell 216), ($4~$m/s, $0\deg$, cell 256) and ($-4~$m/s, $0\deg$, cell 296). We consider the range cell 256 under test. Moreover, the range cells 216 and 296 are kept in the set of secondary data {to illustrate the robustness the shrinkage algorithm in case of contaminated secondary data.} {For both algorithms under study}, four guard cells are removed around the range cell 256. To detect the target and to analyze the performance of the Shrinkage FPE, we use the Adaptive Normalized Matched Filter (ANMF) {introduced by \cite{Conte95} and analyzed in \cite{liu2011acfar, kraut1999cfar, Kraut01, Pascal04-1}. It is} given by:

\begin{equation}\label{anmf}
\Lc(\Mc) = \cfrac{| \Gr p^H \Mc^{-1} \Gr y|^2}{| \Gr p^H \Mc^{-1} \Gr p|\,| \Gr y^H \Mc^{-1} \Gr y|},
\end{equation}
where $\Gr p$ is the so-called STAP steering vector, $\Gr y$ is the observation under test, i.e. in this case the range cell 256 and $\Mc$ is the covariance matrix estimator built on the set of secondary data, i.e. data that are assumed to be signal-free and i.i.d. The ANMF presents some properties of invariance: it is invariant to a multiplicative scale factor on the CM estimator. This implies that there is no need to impose any trace constraint to the other CM estimators. Let us recall that the Shrinkage FPE has, by construction, a trace constraint on its inverse. Now, let us denote $\Lc_{SCM-DL} = \Lc\be\Mc_{SCM-DL}(\beta)\en$, $\Lc_{S-FPE} = \Lc\be\Mc_{S-FPE}(\beta)\en$ and $\Lc_{S-FPE-W} = \Lc\be\Mc_{S-FPE-W}(\beta)\en$, where $$\Mc_{SCM-DL}(\beta) = \cfrac{1-\beta}{N}\disp\sum_{n=1}^N \Gr x_n^H \Gr x_n + \beta \, \Gr I,$$ $\Mc_{S-FPE}(\beta)$ is the unique solution of equation \eqref{FPE-shrinkage} {and $\Mc_{S-FPE-W}(\beta)$ is a solution of equation \eqref{FPE-shrinkage-AW}}. {Moreover, let us denote $\Lc_{SCM}$ the ANMF built with the classical SCM that will play the role of a benchmark.}\\

\begin{figure}[!htp]
\begin{center}
\includegraphics[width=1\linewidth,height=12cm]{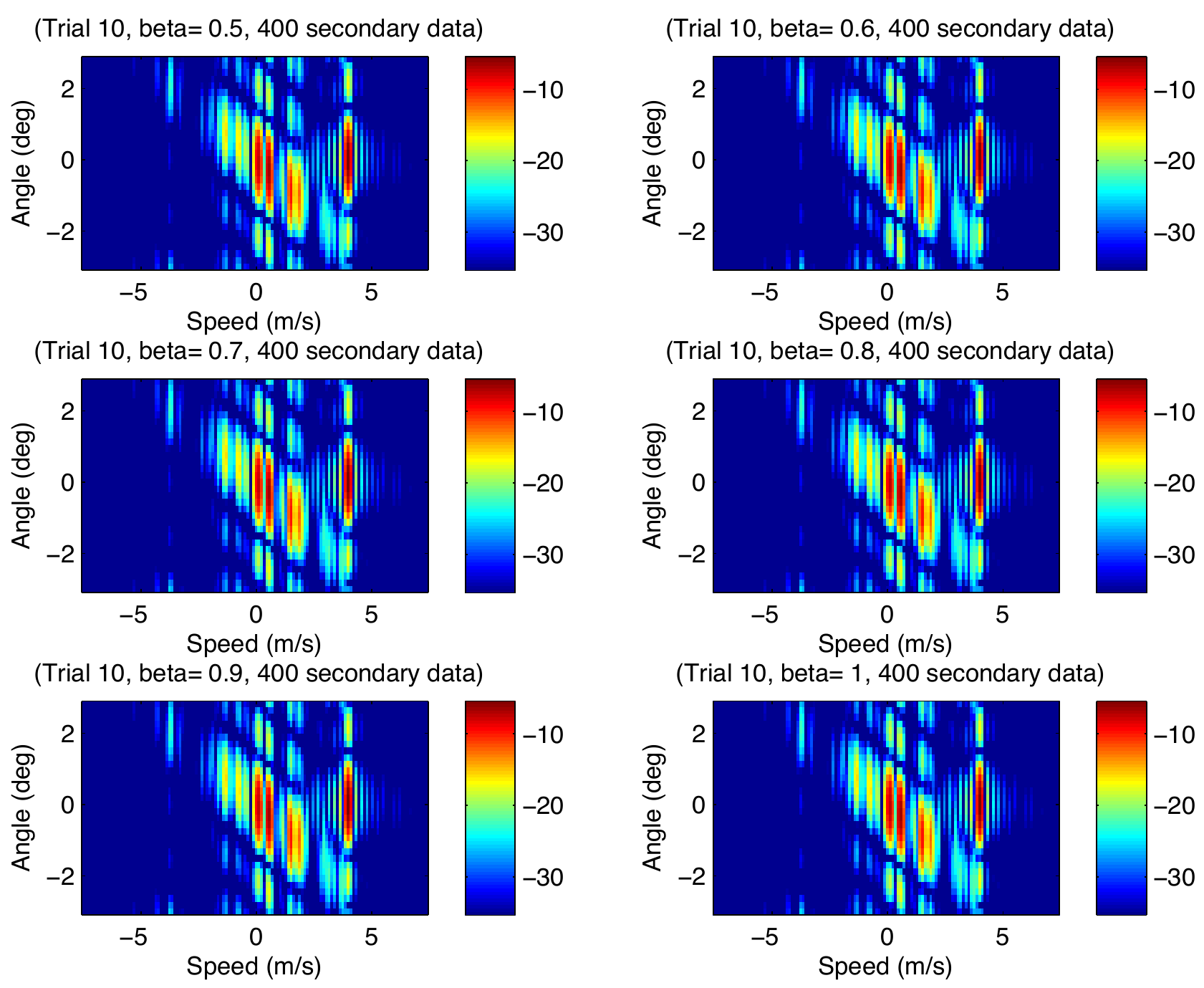}
\caption{$\log_{10}(\Lc_{SCM-DL})$ for different values of $\beta$: angle/speed map with parameters $m=256$, $N=400$. The colour scale is in dB with a dynamic of 30 dB from the maximum value of $\log_{10}(\Lc_{SCM})$}
\label{fig-STAP-SCM-DL-N400}
\end{center}
\end{figure} 

Figure \ref{fig-STAP-SFPE-N400} {(resp. figure \ref{fig-STAP-SFPE-W-N400})} depicts the detection results obtained with $\Lc_{S-FPE}$ {(resp. $\Lc_{S-FPE-W}$)} while on figure \ref{fig-STAP-SCM-DL-N400} {for different speeds and different azimuths}, the results for $\Lc_{SCM-DL}$ are given. More precisely, figures \ref{fig-STAP-SFPE-N400}, \ref{fig-STAP-SFPE-W-N400} and \ref{fig-STAP-SCM-DL-N400} represent the detection results for the range cell 256 in colour scale; the $y$-axis corresponds to the target angle (in degree) while the $x$-axis corresponds for the target velocity (in $m.s^{-1}$). These two figures have been obtained for $N=400$ secondary data, which means that it is a classical over-sampled case. Finally, for each case, results are given for 6 different values of the Shrinkage parameter $\beta$, to highlight the impact of this parameter onto the detection performance.

The first comment is that the well-known diagonal loading techniques allows one to improve the clutter rejection except for the SCM-DL on the clutter ridge. More interestingly, one can notice that the result for the SCM-DL is the same for all values of $\beta$, which means that there is no adaptive whitening (clutter cancellation) with the covariance matrix estimate. The result is the same as when plugging the identity matrix ($\beta=1$) in the ANMF instead of an estimate. The clutter which is mainly on the diagonal is not removed. On the opposite, the S-FPE provides very interesting results since, according to the values of $\beta$, the clutter is more or less totally removed. The cases where $\beta$ equals 0.7 and 0.8 provide the best clutter rejection. {On the other hand, one can notice that the bad results of the SCM-DL are due to the presence of targets in the secondary data while the performance of the S-FPE, due to the robustness of the Tyler's estimator, are not affected by these contaminated data. Finally, the S-FPE and the S-FPE-W seem to provide very similar results (the differences are of order $10^{-5}$) which could be explained by the fact that the resulting covariance matrix estimators are equal up to a scaling constant.} \\ 

\begin{figure}[!htp]
\begin{center}
\includegraphics[width=1\linewidth,height=12cm]{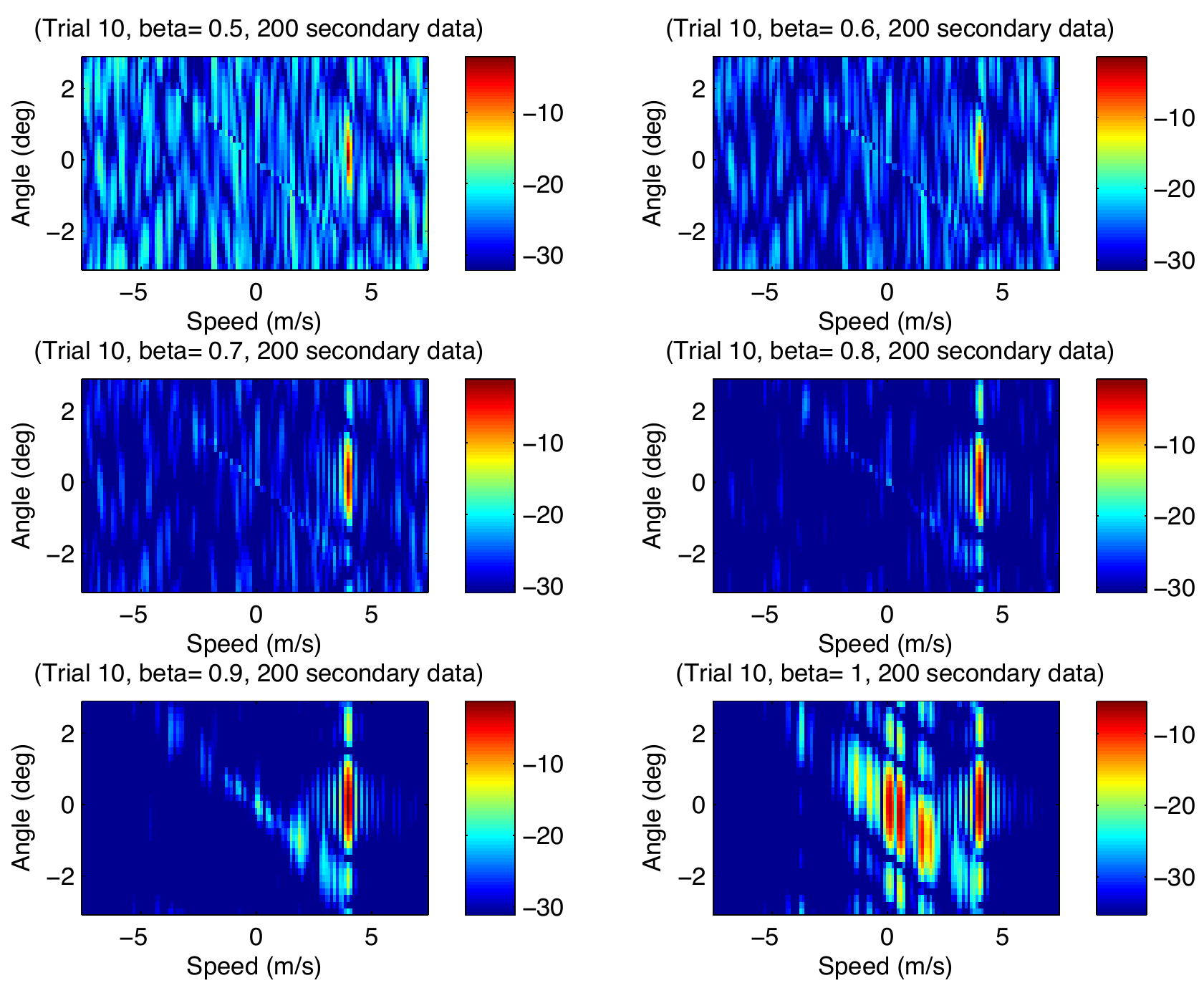}
\caption{$\log_{10}(\Lc_{S-FPE})$ for different values of $\beta$: angle/speed map with parameters $m=256$, $N=200$. The colour scale is in dB with a dynamic of 30 dB from the maximum value of $\log_{10}(\Lc_{S-FPE})$}
\label{fig-STAP-SFPE-N200}
\end{center}
\end{figure} 

\begin{figure}[!htp]
\begin{center}
\includegraphics[width=1\linewidth,height=12cm]{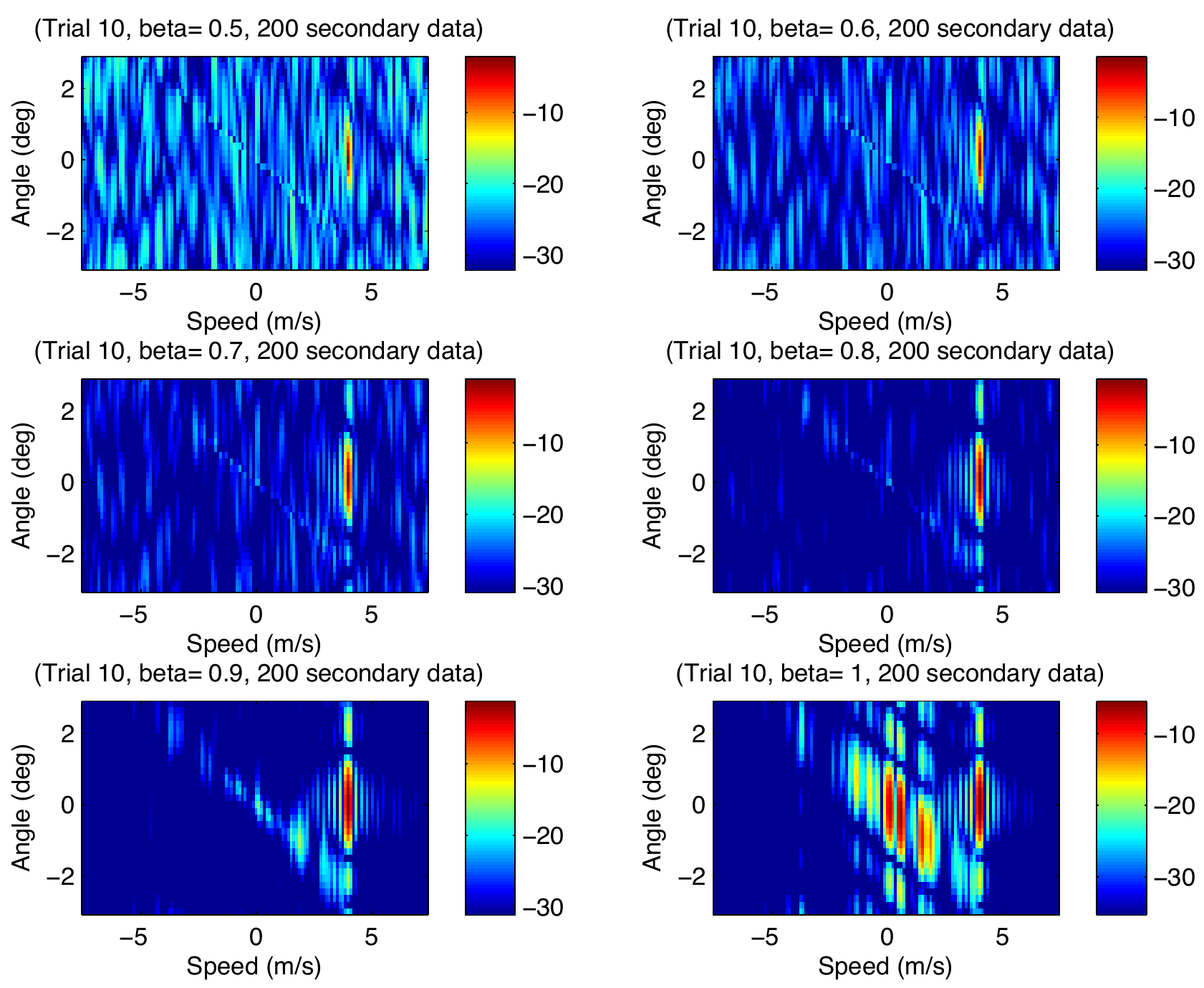}
\caption{$\log_{10}(\Lc_{S-FPE-W})$ for different values of $\beta$: angle/speed map with parameters $m=256$, $N=200$. The colour scale is in dB with a dynamic of 30 dB from the maximum value of $\log_{10}(\Lc_{S-FPE-W})$}
\label{fig-STAP-SFPE-W-N200}
\end{center}
\end{figure} 

Then, since in a STAP context, the ground clutter has been shown to be low-rank \cite{brennan1992subclutter}, it is possible to estimate the covariance matrix with less secondary data by using DL techniques. For this dataset, the estimated clutter rank, obtained from the Brennan rule \cite{brennan1992subclutter} is then equal to $r=45$. This value is small in comparison to the full size of clutter covariance matrix, $m=SM=256$. 

\begin{figure}[!htp]
\begin{center}
\includegraphics[width=1\linewidth,height=12cm]{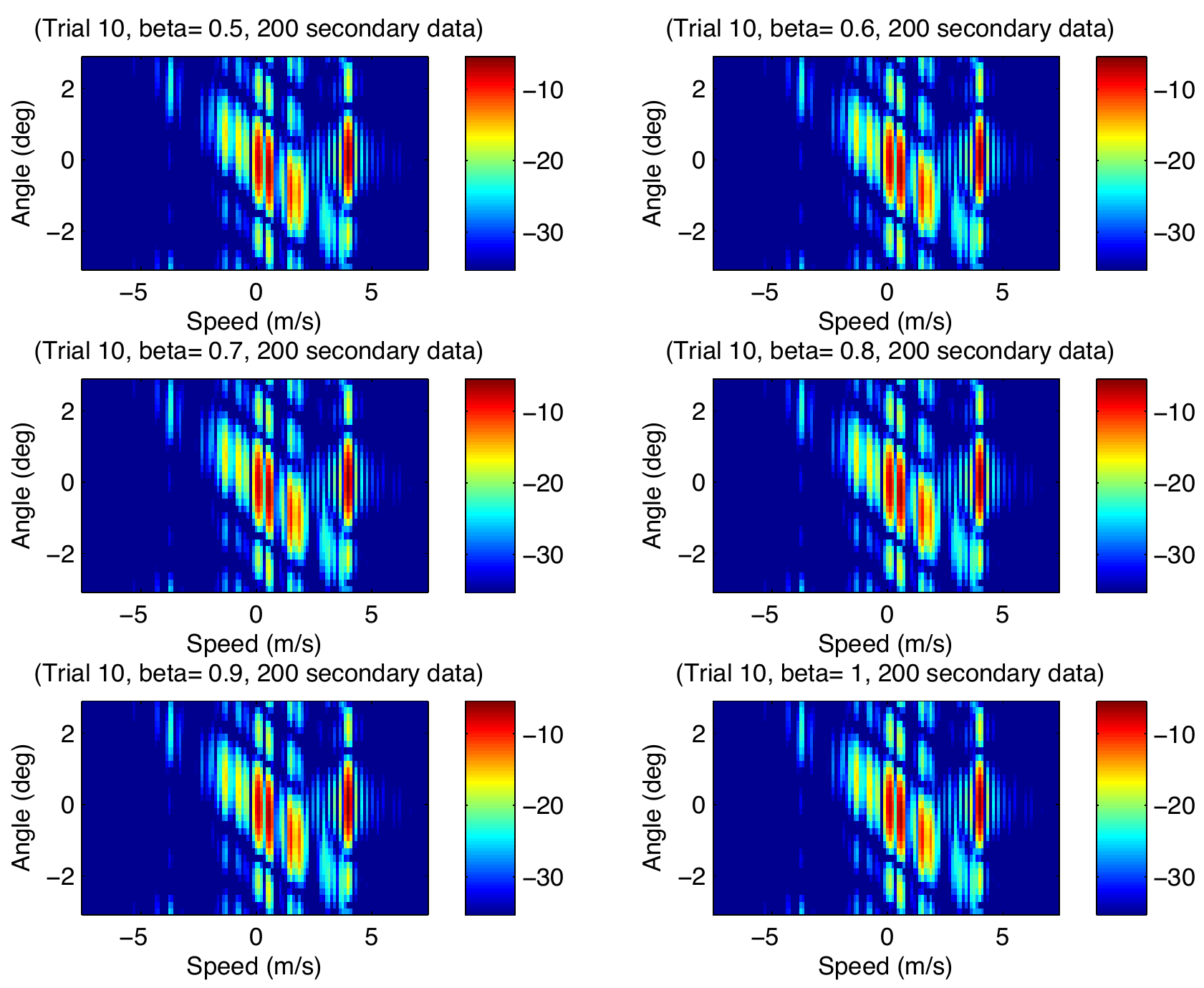}
\caption{$\log_{10}(\Lc_{SCM-DL})$ for different values of $\beta$: angle/speed map with parameters $m=256$, $N=200$. The colour scale is in dB with a dynamic of 30 dB from the maximum value of $\log_{10}(\Lc_{SCM})$}
\label{fig-STAP-SCM-DL-N200}
\end{center}
\end{figure}

This is the purpose of figures \ref{fig-STAP-SFPE-N200}, \ref{fig-STAP-SFPE-W-N200} and \ref{fig-STAP-SCM-DL-N200} where the number of secondary data used to estimate the CM is $N=200$, which is less than the data dimension. Thus, without any DL techniques, the CM estimate is not invertible. First, one can notice that the result for the DL-SCM are very similar as previous ones, due to the fact that the term with the identity matrix is prevailing on the SCM. More importantly, concerning the S-FPE, the DL approach, for $\beta$ large enough, allows to compute the FP algorithm which requires a matrix inversion. Let us recall that, according to Theorem \ref{thm-exist-shrinkage}, $\beta$ has to be greater than $1-N/m$, i.e. approximately 0.22. Moreover, this leads to good results in terms of clutter rejection as well as of target detection. Moreover, the "optimal"\footnote{optimal in the sense that there is a good clutter cancellation} value of $\beta$ has changed and is now closer to 0.8. {Finally,  previous comments concerning the robustness of the S-FPE to the contaminated data are still valid, as well as the similar results between S-FPE and S-FPE-W.}\\

\begin{figure}[!htp]
\begin{center}
\subfigure[N=400]{\includegraphics[width=4cm,height=3cm]{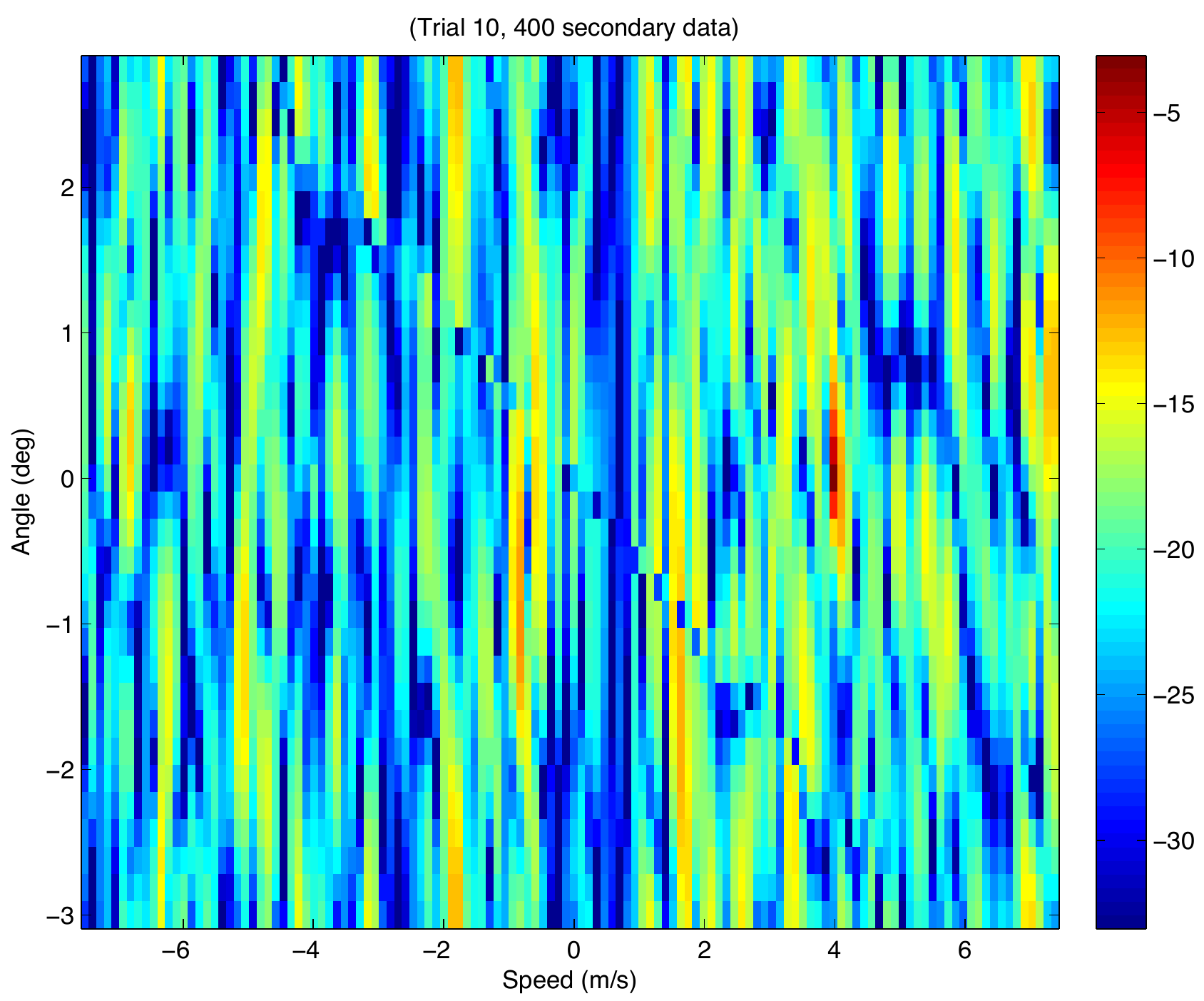}}
\subfigure[N=200]{\includegraphics[width=4cm,height=3cm]{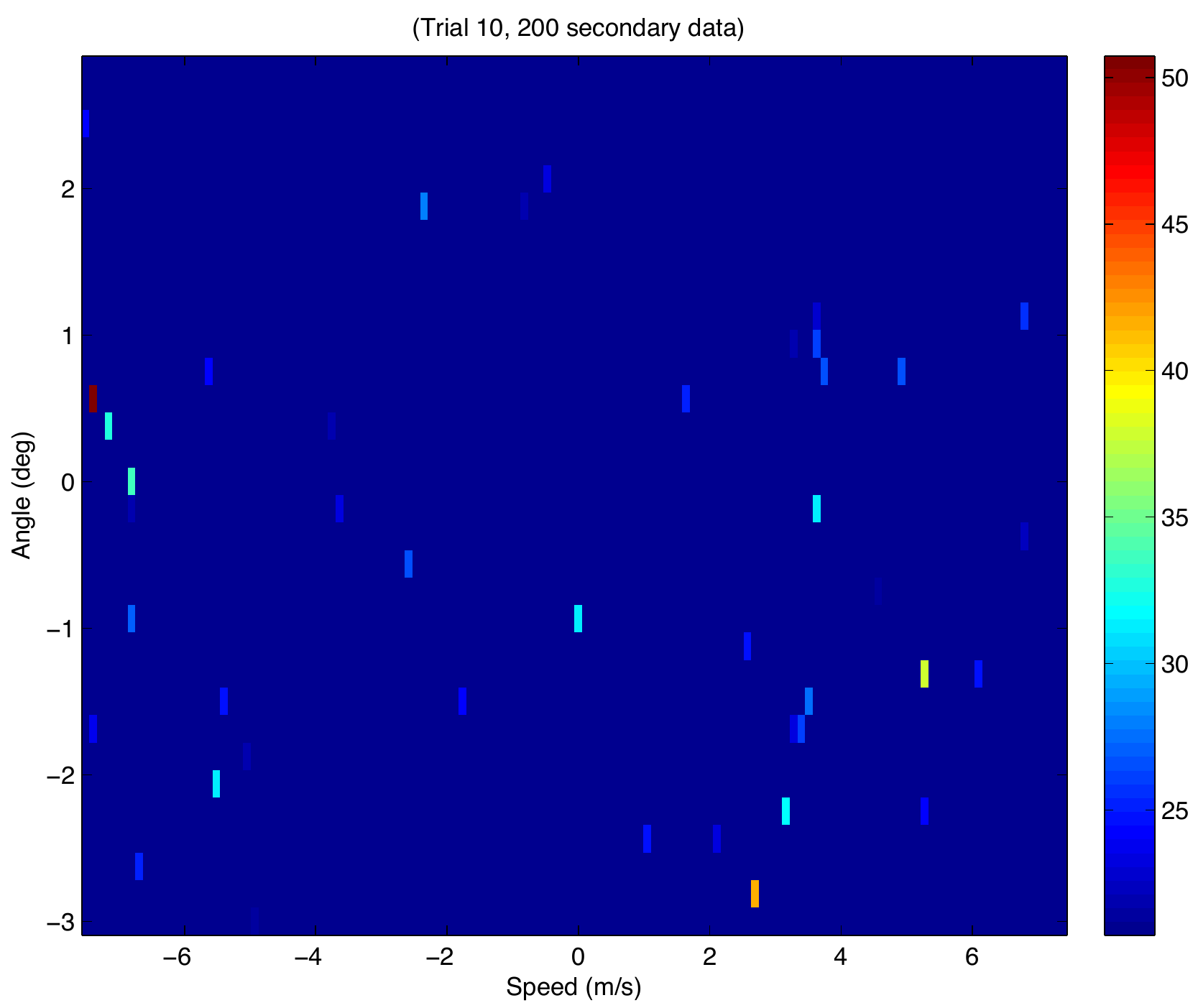}}
\caption{$\log_{10}(\Lc_{SCM})$: angle/speed map with parameters $m=256$ for the range bin 268 and for different values of $N$. The colour scale is in dB with a dynamic of 30 dB from the maximum value of $\log_{10}(\Lc_{SCM})$}
\label{fig-STAP-SCM-256}
\end{center}
\end{figure}

{Now, to highlight the improvement brought by the shrinkage techniques, figure \ref{fig-STAP-SCM-256} depicts the results obtained by the "classical" ANMF built with the SCM, $\Lc_{SCM}$, in the same context as for other detectors, for $N=400$ and $N=200$. As expected, when $N=400$, the performance are degraded in comparison of the performance of other detectors. However, the target can be detected but with a strong clutter level. Then, for a smaller number of secondary data, i.e. $N=200$, $\Lc_{SCM}$ is not able anymore to detect the target .}

\section{Conclusion}
\label{sec:conclu}
In the context of covariance matrix estimation, this paper presents the derivation, namely the proofs of existence and uniqueness, of the shrinkage Fixed Point estimator. Contrary to the results presented in \cite{chen2011robust}, this proof does not require any additional constraint on the trace of the shrinkage FP. However, this more general case has some limitations since it is proved that the existence and uniqueness of the Shrinkage FP are only valid for certain values of the shrinkage parameter. More precisely, the results is true for $m/N<1$ and when $m/N\geq 1$, $\beta$ has to be greater than $1-N/m$, which seems to be a realistic constraint. One the other hand, the performance of the Shrinkage FP has been analyzed through Monte-Carlo simulations, and then, it has been applied on STAP data for target detection purposes. These results show the interest of using such a shrinkage method since it improves the detection performance and presents the main advantage of being able to deal with problems where the number of samples is less than the dimension of the observations.

\appendices
{\section{Proof of Proposition \ref{prop-comparaison}}
\label{app-prop-comparaison}
Define on $\D$ the functional $L$ by
\begin{equation}\label{func-L-1}
L(\Sig)= \cfrac{\exp(-\Tr(\Sig^{-1}))}{\det(\Sig)}.
\end{equation}
Then, one trivially has 
\begin{equation}\label{est0}
F_\beta(\Sigma)=L(\Sig)^{N\beta}F_0(\Sig)^{(1-\beta)}.
\end{equation}
Moreover, note that the real-valued function $x\mapsto x\exp(-x)$ defined on $(\RR^+)^{\ast}$ is upper bounded by one and tends to zero as $x$ tends either to zero or to $+\infty$. Therefore, $L(\Sig)$ tends to zero as soon as one of the eigenvalues of $\Sig$ tends to zero or $+\infty$.
By the properties of $F_0$ proved in Theorem IV.1 of \cite{pascal2008covariance}, $F_0$
is bounded over $\D$ if  $m/N<1$ and one deduces at once 
Items $(a)$ and $(b)$.\\
}

{\section{Proof of Proposition \ref{prop-hessien}}
\label{app-prop-hessien}
By a trivial computation, one has, for every $\beta\in (0,1)$ and symmetric matrix $Q$, that
\begin{eqnarray*}
\frac{Hess_\beta(\Sigma)(Q)}{\beta F_\beta(\Sig)}&=&\langle Q,d\nabla \log F_\beta(\Sig)(Q)\rangle\\
&=&-N\langle Q, \Sig^{-1}V(\Sig,Q,\beta)\Sig^{-1}\rangle,
\end{eqnarray*}
where 
$$
V(\Sig,Q,\beta):=Q-\cfrac{(1-\beta)m}{N}\,\disp \sum_{n=1}^N \cfrac{\xg_n^{\top}  \Sig^{-1} \, Q\Sig^{-1}\xg_n}{(\xg_n^{\top}  \Sig^{-1} \, \xg_n)^2}\xg_n \, \xg_n^{\top}.
$$
Setting {$R:=\Sig^{-1/2}\, Q \, \Sig^{-1/2}$} and using the vectors $\zg_n$ defined in Eq. \eqref{eq00}, one rewrites the above equation as 
$$
\frac{Hess_\beta(\Sigma)(Q)}{N\beta F_\beta(\Sig)}=-\big[\Tr(R^2)-\cfrac{(1-\beta)m}{N}\,\disp \sum_{n=1}^N(\zg_n^{\top}  R \, \zg_n)^2\big].
$$
Multiplying Eq. \eqref{eq00} on the left and on the right by $R$, then taking the trace, one gets an expression for $\Tr(R^2)$ which is reported in the above equation. One therefore obtains that 
\begin{eqnarray*}
\frac{Hess_\beta(\Sigma)(Q)}{N\beta F_\beta(\Sig)}&=&-\big[\beta\Tr(R\Sig^{-1}R)\\
&+&\cfrac{(1-\beta)m}{N}\,\disp \sum_{n=1}^N(\Vert R\zg_n\Vert^2-(\zg_n^{\top}  R \, \zg_n)^2)\big].
\end{eqnarray*}
Since the vectors $\zg_n$ have unit length, one deduces at once from Cauchy-Schwarz's inequality, that the summation term in the above equality is non negative. Hence Eq. \eqref{est1}. 
Therefore, if $\Sig\in \D$ is a critical point of $F_\beta$, $Hess_\beta(\Sigma)$ is negative definite, implying that $\Sig$ is a local strict maximum for $F$.\\
}

{\section{Proof of Proposition \ref{prop-LB}}
\label{app-prop-LB}
Let $P$ be the unique fixed point of $F_0$ such that $\Tr(P)=m$. 
Then, for every $\beta\in (0,1)$, one has $F_\beta(\Sig(\beta))\geq F_\beta(P)$ and 
 $F_0(\Sig(\beta))\leq F_0(P)$. We multiply the two inequalities and, after using Eq. \eqref{est0} and the fact that $\Tr(\Sig(\beta)^{-1})=m$, we deduce that, for every $\beta\in (0,1)$,
 $$
 \det(\Sig(\beta))\leq \frac{\exp(-m)}{L(P)}.\\
 $$
}

{\section{Proof of Proposition \ref{prop-ext}}
\label{app-prop-ext}
Set $\gamma:=(1-\beta)m/N$ and assume first that Eq. \eqref{fpe-1} admits a solution $\Sig$ in $\D$. Set $d_1:=\xg_1^T\Sig^{-1}\xg_1$ and $\Sig_1:=\Sig-\frac{\gamma}{d_1}\xg_1\xg_1^T$. Since $\Sig_1\geq \beta \Ig$, one gets that $\Sig_1$ is positive definite and $\bar{d}_1:=\xg_1^T\Sig_1^{-1}\xg_1$ is strictly positive. By a standard computation, one gets that $d_1=(1-\gamma)\bar{d}_1$, which implies that $\gamma<1$. \\\\
Conversely, assume that $\gamma<1$. A careful examination of the proof of Theorem \ref{thm-exist-shrinkage} reveals that the assumption $m/N<1$ is required only to get that the functional $F_0$ can be extended by continuity on the boundary of $\D$ by zero and bounded over $\D$. It turns out that, under the relaxed condition $\gamma<1$, one can show that 
$F_\beta$ can be extended by continuity on the boundary of $\D$ by zero and $F_\beta(\Sig)$ tends to zero as $\Vert \Sig\Vert$ tends to $+\infty$. To proceed, one can also suppose without loss of generality that $N\leq m$ and the $(x_i)_{1\leq i\leq N}$ are linearly independent. For $\Sig\in\D$, set $G_\beta(\Sig):=F_\beta(\Sig)^{1/N}$. 
If $0<\lambda_1(\Sig)\leq \cdots\leq \lambda_m(\Sig)$ are used to denote the $m$-th ordered eigenvalues of $\Sig$, we define on $\D$ the following functional for $\rho>0$
\begin{multline}\label{toz1}
H_\rho(\Sig):=\Big(\disp  \prod_{n=1}^N \frac{\exp(-\rho/\lambda_n(\Sig))}{\lambda_n(\Sig)^{1-\gamma}}\Big)\cdot\\ \Big(\disp  \prod_{n=N+1}^m \frac{\exp(-\rho/\lambda_n(\Sig))}{\lambda_n(\Sig)}\Big).
\end{multline}
Moreover, complete orthogonally the $(x_i)_{1\leq i\leq N}$ by an orthonormal set $(y_k)_{N+1\leq i\leq m}$ and let $Z$ be the $m\times m$ matrix whose columns are the $(x_i)_{1\leq i\leq N}$ and then the $(y_k)_{N+1\leq i\leq m}$. Finally note $\widetilde\Sig:=Z^{-1}\Sig(Z^{-1})^T$ for every $\Sig\in\D$. 
Then one has 
$$
G_\beta(\Sig)=\frac{\exp(-\beta\Tr((Z^TZ)^{-1}\widetilde\Sig)}{\det(Z^{-2})\det(\widetilde\Sig)}\disp  \prod_{n=1}^N
\Big(\widetilde\Sig^{-1}\Big)_{nn}^{-\gamma}.
$$
We finally need the following lemma before estimating $G_\beta$.
\begin{Lem}
\begin{description} 
\item[$(i)$] If $A,B$ are two symmetric positive definite $m\times m$ matrices, then there exist two positive constants $a_1,a_2$ only depending on $A$ such that $a_1\Tr(B)\leq \Tr(AB)\leq a_2 \Tr(B)$.
\item[$(ii)$] If $A$ is a  symmetric positive definite $m\times m$ matrix, then
$$
\disp  \prod_{j=1}^N A_{jj}\geq \disp  \prod_{j=1}^N \lambda_j(A).
$$
\end{description}
\end{Lem}
\begin{IEEEproof}
The right inequality in Item $(i)$ follows with from Von Neuman's theorem (cf. Theorem $7.4.11$ in \cite{horn2012matrix})
and the the left inequality as a application of the right inequality with $A^{-1}$ and $A^{1/2}BA^{1/2}$ instead of $A$ and $B$.
As for Item $(ii)$, first consider $A_N=(A_{ij})_{1\leq i,j\leq N}$. According to Hadamard's inequality, one gets
$\disp  \prod_{j=1}^N A_{jj}\geq \det(A_N)$. Then one concludes by using Theorem $4.3.28$  in \cite{horn2012matrix} stating that $\lambda_j(A_N)\geq \lambda_j(A)$ for $1\leq j\leq N$. \\\\
\end{IEEEproof}
Now, we deduce at once from the previous lemma and Eq. \eqref{toz1} that
$G_\beta(\Sig)\leq C_0H_{a\beta}(\widetilde\Sig)$, where the constants $C_0$ and $a$ only depend on $Z$.\\\\
Note that the real-valued function $x\mapsto x\exp(-a\beta x)$ defined on $(\RR^+)^{\ast}$ is upper bounded and tends to zero as $x$ tends either to zero or to $+\infty$. The same result holds true for $x\mapsto x^{1-\gamma}$ since $\gamma<1$. One gets immediately that $H_{a\beta}$ (and then $G_\beta$) admits a global maximum on $\D$ and one concludes the proof of Proposition \ref{prop-ext} by using the rest of the argument of Theorem \ref{thm-exist-shrinkage}.\\
}

{\section{Analysis of the global maximum}}
\label{app-prop-maxglob}
In this section, we investigate the more general question of maximizing $F$ over the full state space $[0,1]\times \D$. Taking into account the previous results, it is enough to maximize over $[0,1]$ the function $M(\beta):=\log F(\beta,\Sig(\beta))$. We get the following.
\begin{Prop}
\label{prop-maxglob}
The maximum of  $F$ over $[0,1]\times \D$ is reached for $\beta$ equal to zero or one and, therefore, the set of maximizers is either $Id_m$ or the half-line $\mathrm{R}_*^+P$, where $P$ is the unique fixed point of $F_0$ such that $\Tr(P)=m$. 
\end{Prop}
\begin{IEEEproof} 
We will derivate the function $M(\cdot)$ and for that, we next prove that $M(\cdot)$ is of class $C^2$  over $(0,1]$. To show that, it amounts to prove that the function $\beta\mapsto \Sig(\beta)$ is itself of class $C^2$ over $(0,1]$. To see the latter, notice that $\Sig(\beta)$ is defined implicitly by the equation 
\begin{equation}\label{est11}
\frac{\partial \log F}{\partial \Sig}(\beta,\Sig(\beta))=0,
\end{equation}
for $\beta\in (0,1]$. Since $F$ is real-analytic with respect to its arguments, if one is able to apply the implicit function theorem to the above equation, then at once one gets the desired regularity for $\beta\mapsto \Sig(\beta)$ over  $(0,1]$. In turn, to meet the conditions of the implicit function theorem, one must get the invertibility of the map $\frac{\partial^2 \log F}{\partial \Sig^2}(\beta,\Sig(\beta))$, i.e. that of  $Hess_\beta(\Sig(\beta))$, which is established in Proposition \ref{prop-hessien}.

For $\beta\in (0,1]$, one has Eq. \eqref{est11} and one deduces that 
\begin{eqnarray}
M'(\beta)&=&\frac{\partial \log F}{\partial \beta}(\beta,\Sig(\beta))\\
&=& -Nm +m\disp\sum_{n=1}^N \log(\xg_n^{\top}  \Sig^{-1}(\beta) \xg_n). \label{est22}
\end{eqnarray}
We next prove that $M''(\beta)>0$ for $\beta\in (0,1]$. First, taking the derivative with respect to $\beta$ in Eqs. \eqref{est11} and \eqref{est22} yields that, for $\beta\in (0,1]$,
$$
\frac{\partial^2 \log F}{\partial \beta\partial \Sig}(\beta,\Sig(\beta))+\frac{\partial^2 \log F}{\partial \Sig^2}(\Sig'(\beta),\cdot)=0,
$$
and on the other hand
$$
M''(\beta)=\frac{\partial^2 \log F}{\partial \beta\partial \Sig}(\beta,\Sig(\beta))(\Sig'(\beta).
$$
One immediately deduces that, for $\beta\in (0,1]$,
$$
M''(\beta)=-Hess_\beta(\Sig'(\beta))>0.
$$
We deduce that $M$ is convex on $[0,1]$ and, therefore, achieves its maximum at $\beta$ equal to zero or one.
\end{IEEEproof}
\begin{rem}
According to the above proposition, one must compare one with $\disp  \prod_{n=1}^N  (\xg_n^{\top}  \Sig^{-1} \xg_n)$ to decide whether the maximum is reached for $\beta=0$ or $\beta=1$.
\end{rem}

\section*{Aknowledgement}
The authors would like to thank Dr. Romain Couillet for pointing out the necessary part of Proposition \ref{prop-ext}. 
\bibliographystyle{IEEEtran} 
\bibliography{biblio}

\end{document}

%% file: definitions.tex
\def\Gr#1{\mathbf #1}
\def\RR{\mathbb{R}}

\def\D{\mathcal{D}}

\def\CN{\mathcal{CN}}

\def\ag{\mathbf{a}}

\def\xg{\mathbf{x}}

\def\zg{\mathbf{z}}

\def\Lc{\widehat{\Lambda}}

\def\Sig{\boldsymbol{\Sigma}}

\def\Ig{\mathbf{I}}

\def\Mg{\mathbf{M}}

\newcommand\Mc{\widehat{\Gr{M}}}

\newtheorem{Thm}{Theorem}[section]
\newtheorem{Lem}{Lemma}[section]

\newtheorem{Prop}{Proposition}[section]
\newtheorem{rem}{Remark}[section]

\newcommand\be{\left(}

\newcommand\en{\right)}

\newcommand\disp{\displaystyle}

\def\Tr{\,\textup{Tr}}

%% file: fig1.tex
\begin{figure}[!h]
\begin{center}
\subfigure[$\rho=0.01$]{\label{err-rho=0.01}\begin{tikzpicture}[font=\footnotesize,scale=0.7]
\pgfplotsset{every axis/.append style={mark options=solid, mark size=2.5pt}}
\pgfplotsset{every axis legend/.append style={fill=white,cells={anchor=west},at={(0.98,0.68)},anchor=north east}} \tikzstyle{every axis y label}+=[yshift=-10pt]
\tikzstyle{every axis x label}+=[yshift=5pt]
\tikzstyle{dashed dotted}=[dash pattern=on 1pt off 4pt on 6pt off 4pt]

\begin{axis}[xlabel={$\beta$},ylabel={NMSE},
xmin=0,xmax=1,ymin=0,ymax=2.25
]
\addplot[mark=star,smooth,blue,line width=.5pt] plot coordinates {
(0.001,2.00634)(0.005,2.00634)(0.01,2.00634)(0.05,2.00634)(0.1,2.00634)(0.15,2.00634)(0.2,2.00634)(0.25,2.00634)(0.3,2.00634)(0.35,2.00634)(0.4,2.00634)(0.45,2.00634)(0.5,2.00634)(0.55,2.00634)(0.6,2.00634)(0.65,2.00634)(0.7,2.00634)(0.75,2.00634)(0.8,2.00634)(0.85,2.00634)(0.9,2.00634)(0.95,2.00634)(1,2.00634)
};
\addplot[mark=triangle,smooth,red,line width=.5pt] plot coordinates {
(0.001,2.00031)(0.005,1.97644)(0.01,1.94714)(0.05,1.73244)(0.1,1.50456)(0.15,1.31135)(0.2,1.14523)(0.25,1.00087)(0.3,0.874351)(0.35,0.762691)(0.4,0.663562)(0.45,0.575097)(0.5,0.495775)(0.55,0.424333)(0.6,0.359708)(0.65,0.300995)(0.7,0.247417)(0.75,0.198304)(0.8,0.153082)(0.85,0.111271)(0.9,0.0725408)(0.95,0.0370969)(1,0.0135407)
};
\legend{ {FPE},{Shrinkage FPE}};
\end{axis}
\end{tikzpicture}
}

\subfigure[$\rho=0.5$]{\label{err-rho=0.5}\begin{tikzpicture}[font=\footnotesize,scale=0.7]
\pgfplotsset{every axis/.append style={mark options=solid, mark size=2.5pt}}
\pgfplotsset{every axis legend/.append style={fill=white,cells={anchor=west},at={(0.98,0.98)},anchor=north east}} \tikzstyle{every axis y label}+=[yshift=-10pt]
\tikzstyle{every axis x label}+=[yshift=5pt]
\tikzstyle{dashed dotted}=[dash pattern=on 1pt off 4pt on 6pt off 4pt]

\begin{axis}[xlabel={$\beta$},ylabel={NMSE},
xmin=0,xmax=1,ymin=0,ymax=2.25
]
\addplot[mark=star,smooth,blue,line width=.5pt] plot coordinates {
(0.001,1.75665)(0.005,1.75665)(0.01,1.75665)(0.05,1.75665)(0.1,1.75665)(0.15,1.75665)(0.2,1.75665)(0.25,1.75665)(0.3,1.75665)(0.35,1.75665)(0.4,1.75665)(0.45,1.75665)(0.5,1.75665)(0.55,1.75665)(0.6,1.75665)(0.65,1.75665)(0.7,1.75665)(0.75,1.75665)(0.8,1.75665)(0.85,1.75665)(0.9,1.75665)(0.95,1.75665)(1,1.75665)
};
\addplot[mark=triangle,smooth,red,line width=.5pt] plot coordinates {
(0.001,1.74968)(0.005,1.72213)(0.01,1.68839)(0.05,1.44348)(0.1,1.18925)(0.15,0.980753)(0.2,0.810355)(0.25,0.67383)(0.3,0.569134)(0.35,0.495297)(0.4,0.45105)(0.45,0.433404)(0.5,0.437058)(0.55,0.455302)(0.6,0.481796)(0.65,0.511787)(0.7,0.542244)(0.75,0.571421)(0.8,0.5984)(0.85,0.622768)(0.9,0.644422)(0.95,0.663443)(1,0.680025)
};
\legend{ {FPE},{Shrinkage FPE}};
\end{axis}
\end{tikzpicture}
}

\subfigure[$\rho=0.99$]{\label{err-rho=0.99}\begin{tikzpicture}[font=\footnotesize,scale=0.7]
\pgfplotsset{every axis/.append style={mark options=solid, mark size=2.5pt}}
\pgfplotsset{every axis legend/.append style={fill=white,cells={anchor=west},at={(0.98,0.98)},anchor=north east}} \tikzstyle{every axis y label}+=[yshift=-10pt]
\tikzstyle{every axis x label}+=[yshift=5pt]
\tikzstyle{dashed dotted}=[dash pattern=on 1pt off 4pt on 6pt off 4pt]

\begin{axis}[xlabel={$\beta$},ylabel={NMSE},
xmin=0,xmax=1,ymin=0,ymax=2.25
]
\addplot[mark=star,smooth,blue,line width=.5pt] plot coordinates {
(0.001,1.20707)(0.005,1.20707)(0.01,1.20707)(0.05,1.20707)(0.1,1.20707)(0.15,1.20707)(0.2,1.20707)(0.25,1.20707)(0.3,1.20707)(0.35,1.20707)(0.4,1.20707)(0.45,1.20707)(0.5,1.20707)(0.55,1.20707)(0.6,1.20707)(0.65,1.20707)(0.7,1.20707)(0.75,1.20707)(0.8,1.20707)(0.85,1.20707)(0.9,1.20707)(0.95,1.20707)(1,1.20707)
};
\addplot[mark=triangle,smooth,red,line width=.5pt] plot coordinates {
(0.001,1.20005)(0.005,1.17223)(0.01,1.13809)(0.05,0.888234)(0.1,0.627119)(0.15,0.421162)(0.2,0.275318)(0.25,0.205866)(0.3,0.225165)(0.35,0.312133)(0.4,0.424379)(0.45,0.53015)(0.5,0.623795)(0.55,0.705766)(0.6,0.776732)(0.65,0.837246)(0.7,0.887727)(0.75,0.928484)(0.8,0.959769)(0.85,0.981777)(0.9,0.994257)(0.95,0.998146)(1,0.999018)
};
\legend{ {FPE},{Shrinkage FPE}};
\end{axis}
\end{tikzpicture}
}

\caption{\label{fig-1} \tred{Normalized mean square error} of  $\hat{\Gr M}$ estimated by the FPE and the shrinkage FPE versus the parameter $\beta$ of the shrinkage FPE and for different covariance matrices, i.e. for different values of the correlation coefficient $\rho$, where $N=24$ samples and the dimension of the data is $m=12$ for Gaussian noise.}
\end{center}
\end{figure}

%% file: fig2.tex
\begin{figure}[!h]
\begin{center}
\subfigure[$\rho=0.01$]{\label{err-rho=0.01-N=48}\begin{tikzpicture}[font=\footnotesize,scale=0.7]
\pgfplotsset{every axis/.append style={mark options=solid, mark size=2.5pt}}
\pgfplotsset{every axis legend/.append style={fill=white,cells={anchor=west},at={(0.98,0.98)},anchor=north east}} \tikzstyle{every axis y label}+=[yshift=-10pt]
\tikzstyle{every axis x label}+=[yshift=5pt]
\tikzstyle{dashed dotted}=[dash pattern=on 1pt off 4pt on 6pt off 4pt]

\begin{axis}[xlabel={$\beta$},ylabel={NMSE},
xmin=0,xmax=1,ymin=0,ymax=2.25
]
\addplot[mark=star,smooth,blue,line width=.5pt] plot coordinates {
(0.001,0.802911)(0.005,0.802911)(0.01,0.802911)(0.05,0.802911)(0.1,0.802911)(0.15,0.802911)(0.2,0.802911)(0.25,0.802911)(0.3,0.802911)(0.35,0.802911)(0.4,0.802911)(0.45,0.802911)(0.5,0.802911)(0.55,0.802911)(0.6,0.802911)(0.65,0.802911)(0.7,0.802911)(0.75,0.802911)(0.8,0.802911)(0.85,0.802911)(0.9,0.802911)(0.95,0.802911)(1,0.802911)
};
\addplot[mark=triangle,smooth,red,line width=.5pt] plot coordinates {
(0.001,0.801482)(0.005,0.795787)(0.01,0.788722)(0.05,0.734205)(0.1,0.670673)(0.15,0.611701)(0.2,0.556781)(0.25,0.505489)(0.3,0.457464)(0.35,0.412392)(0.4,0.369998)(0.45,0.330041)(0.5,0.292302)(0.55,0.256586)(0.6,0.222718)(0.65,0.190536)(0.7,0.159898)(0.75,0.130677)(0.8,0.10277)(0.85,0.0761177)(0.9,0.0507874)(0.95,0.0274714)(1,0.0135407)
};
\legend{ {FPE},{Shrinkage FPE}};
\end{axis}
\end{tikzpicture}
}

\subfigure[$\rho=0.5$]{\label{err-rho=0.5-N=48}\begin{tikzpicture}[font=\footnotesize,scale=0.7]
\pgfplotsset{every axis/.append style={mark options=solid, mark size=2.5pt}}
\pgfplotsset{every axis legend/.append style={fill=white,cells={anchor=west},at={(0.98,0.98)},anchor=north east}} \tikzstyle{every axis y label}+=[yshift=-10pt]
\tikzstyle{every axis x label}+=[yshift=5pt]
\tikzstyle{dashed dotted}=[dash pattern=on 1pt off 4pt on 6pt off 4pt]

\begin{axis}[xlabel={$\beta$},ylabel={NMSE},
xmin=0,xmax=1,ymin=0,ymax=1
]
\addplot[mark=star,smooth,blue,line width=.5pt] plot coordinates {
(0.001,0.674456)(0.005,0.674456)(0.01,0.674456)(0.05,0.674456)(0.1,0.674456)(0.15,0.674456)(0.2,0.674456)(0.25,0.674456)(0.3,0.674456)(0.35,0.674456)(0.4,0.674456)(0.45,0.674456)(0.5,0.674456)(0.55,0.674456)(0.6,0.674456)(0.65,0.674456)(0.7,0.674456)(0.75,0.674456)(0.8,0.674456)(0.85,0.674456)(0.9,0.674456)(0.95,0.674456)(1,0.674456)
};
\addplot[mark=triangle,smooth,red,line width=.5pt] plot coordinates {
(0.001,0.672415)(0.005,0.66431)(0.01,0.654309)(0.05,0.579516)(0.1,0.499473)(0.15,0.435413)(0.2,0.388714)(0.25,0.360369)(0.3,0.350062)(0.35,0.355585)(0.4,0.373189)(0.45,0.398709)(0.5,0.428577)(0.55,0.4602)(0.6,0.491864)(0.65,0.522494)(0.7,0.551445)(0.75,0.578354)(0.8,0.603049)(0.85,0.625486)(0.9,0.645713)(0.95,0.663841)(1,0.680025)
};
\legend{ {FPE},{Shrinkage FPE}};
\end{axis}
\end{tikzpicture}
}

\subfigure[$\rho=0.99$]{\label{err-rho=0.99-N=48}\begin{tikzpicture}[font=\footnotesize,scale=0.7]
\pgfplotsset{every axis/.append style={mark options=solid, mark size=2.5pt}}
\pgfplotsset{every axis legend/.append style={fill=white,cells={anchor=west},at={(0.02,0.98)},anchor=north west}} \tikzstyle{every axis y label}+=[yshift=-10pt]
\tikzstyle{every axis x label}+=[yshift=5pt]
\tikzstyle{dashed dotted}=[dash pattern=on 1pt off 4pt on 6pt off 4pt]

\begin{axis}[xlabel={$\beta$},ylabel={NMSE},
xmin=0,xmax=1,ymin=0,ymax=1
]
\addplot[mark=star,smooth,blue,line width=.5pt] plot coordinates {
(0.001,0.395918)(0.005,0.395918)(0.01,0.395918)(0.05,0.395918)(0.1,0.395918)(0.15,0.395918)(0.2,0.395918)(0.25,0.395918)(0.3,0.395918)(0.35,0.395918)(0.4,0.395918)(0.45,0.395918)(0.5,0.395918)(0.55,0.395918)(0.6,0.395918)(0.65,0.395918)(0.7,0.395918)(0.75,0.395918)(0.8,0.395918)(0.85,0.395918)(0.9,0.395918)(0.95,0.395918)(1,0.395918)
};
\addplot[mark=triangle,smooth,red,line width=.5pt] plot coordinates {
(0.001,0.393151)(0.005,0.382174)(0.01,0.368657)(0.05,0.269903)(0.1,0.17593)(0.15,0.140756)(0.2,0.171193)(0.25,0.253069)(0.3,0.347282)(0.35,0.439733)(0.4,0.52581)(0.45,0.605067)(0.5,0.677478)(0.55,0.742998)(0.6,0.801537)(0.65,0.85295)(0.7,0.897041)(0.75,0.933562)(0.8,0.962243)(0.85,0.982771)(0.9,0.994464)(0.95,0.998156)(1,0.999018)
};
\legend{ {FPE},{Shrinkage FPE}};
\end{axis}
\end{tikzpicture}
}

\caption{\label{fig-2} Normalized mean square error of  $\hat{\Gr M}$ estimated by the FPE and the shrinkage FPE versus the parameter $\beta$ of the shrinkage FPE and for different covariance matrices, i.e. for different values of the correlation coefficient $\rho$, where $N=48$ samples and the dimension of the data is $m=12$ for Gaussian noise.}
\end{center}
\end{figure}

%% file: fig2-b.tex
\begin{figure}[!h]
\begin{center}
\subfigure[$\rho=0.5$]{\label{err-rho=0.5-N=200}\begin{tikzpicture}[font=\footnotesize,scale=0.7]
\pgfplotsset{every axis/.append style={mark options=solid, mark size=2.5pt}}
\pgfplotsset{every axis legend/.append style={fill=white,cells={anchor=west},at={(0.98,0.98)},anchor=north east}} \tikzstyle{every axis y label}+=[yshift=-10pt]
\tikzstyle{every axis x label}+=[yshift=5pt]
\tikzstyle{dashed dotted}=[dash pattern=on 1pt off 4pt on 6pt off 4pt]

\begin{axis}[xlabel={$\beta$},ylabel={NMSE},
xmin=0,xmax=1,ymin=0,ymax=1
]
\addplot[mark=star,smooth,blue,line width=.5pt] plot coordinates {
(0.001,0.22399)(0.005,0.22399)(0.01,0.22399)(0.05,0.22399)(0.1,0.22399)(0.15,0.22399)(0.2,0.22399)(0.25,0.22399)(0.3,0.22399)(0.35,0.22399)(0.4,0.22399)(0.45,0.22399)(0.5,0.22399)(0.55,0.22399)(0.6,0.22399)(0.65,0.22399)(0.7,0.22399)(0.75,0.22399)(0.8,0.22399)(0.85,0.22399)(0.9,0.22399)(0.95,0.22399)(1,0.22399)
};
\addplot[mark=triangle,smooth,red,line width=.5pt] plot coordinates {
(0.001,0.223314)(0.005,0.220667)(0.01,0.217488)(0.05,0.197695)(0.1,0.188531)(0.15,0.196781)(0.2,0.218985)(0.25,0.250001)(0.3,0.285532)(0.35,0.322833)(0.4,0.3603)(0.45,0.396998)(0.5,0.432364)(0.55,0.466054)(0.6,0.497856)(0.65,0.527648)(0.7,0.555369)(0.75,0.581008)(0.8,0.604594)(0.85,0.626188)(0.9,0.645879)(0.95,0.663781)(1,0.680025)
};
\legend{ {FPE},{Shrinkage FPE}};
\end{axis}
\end{tikzpicture}
}

\subfigure[$\rho=0.99$]{\label{err-rho=0.99-N=200}\begin{tikzpicture}[font=\footnotesize,scale=0.7]
\pgfplotsset{every axis/.append style={mark options=solid, mark size=2.5pt}}
\pgfplotsset{every axis legend/.append style={fill=white,cells={anchor=west},at={(0.02,0.98)},anchor=north west}} \tikzstyle{every axis y label}+=[yshift=-10pt]
\tikzstyle{every axis x label}+=[yshift=5pt]
\tikzstyle{dashed dotted}=[dash pattern=on 1pt off 4pt on 6pt off 4pt]

\begin{axis}[xlabel={$\beta$},ylabel={NMSE},
xmin=0,xmax=1,ymin=0,ymax=1
]
\addplot[mark=star,smooth,blue,line width=.5pt] plot coordinates {
(0.001,0.0856946)(0.005,0.0856946)(0.01,0.0856946)(0.05,0.0856946)(0.1,0.0856946)(0.15,0.0856946)(0.2,0.0856946)(0.25,0.0856946)(0.3,0.0856946)(0.35,0.0856946)(0.4,0.0856946)(0.45,0.0856946)(0.5,0.0856946)(0.55,0.0856946)(0.6,0.0856946)(0.65,0.0856946)(0.7,0.0856946)(0.75,0.0856946)(0.8,0.0856946)(0.85,0.0856946)(0.9,0.0856946)(0.95,0.0856946)(1,0.0856946)
};
\addplot[mark=triangle,smooth,red,line width=.5pt] plot coordinates {
(0.001,0.0848764)(0.005,0.0818125)(0.01,0.0784498)(0.05,0.0769055)(0.1,0.134859)(0.15,0.218842)(0.2,0.301068)(0.25,0.379947)(0.3,0.455273)(0.35,0.526885)(0.4,0.594614)(0.45,0.658263)(0.5,0.717602)(0.55,0.772362)(0.6,0.822228)(0.65,0.866833)(0.7,0.905754)(0.75,0.938521)(0.8,0.96465)(0.85,0.983625)(0.9,0.994598)(0.95,0.998161)(1,0.999018)
};
\legend{ {FPE},{Shrinkage FPE}};
\end{axis}
\end{tikzpicture}
}

\caption{\label{fig-2-N=200} Normalized mean square error of  $\hat{\Gr M}$ estimated by the FPE and the shrinkage FPE versus the parameter $\beta$ of the shrinkage FPE and for different covariance matrices, i.e. for different values of the correlation coefficient $\rho$, where $N=200$ samples and the dimension of the data is $m=12$ for Gaussian noise.}
\end{center}
\end{figure}

%% file: fig3-cv.tex
\begin{figure}[!h]
\begin{center}
{\label{err-cv=0.99}\begin{tikzpicture}[font=\footnotesize,scale=0.8]
\pgfplotsset{every axis/.append style={mark options=solid, mark size=2.5pt}}
\pgfplotsset{every axis legend/.append style={fill=white,cells={anchor=west},at={(0.02,0.98)},anchor=north west}} \tikzstyle{every axis y label}+=[yshift=-10pt]
\tikzstyle{every axis x label}+=[yshift=5pt]
\tikzstyle{dashed dotted}=[dash pattern=on 1pt off 4pt on 6pt off 4pt]

\begin{axis}[xlabel={$\beta$},ylabel={$C_1(\beta)$}, xmin=0,xmax=1,ymin=0,ymax=1]
\addplot[mark=star,smooth,green,line width=.5pt] plot coordinates {
(0.001,0.00296084)(0.005,0.0147376)(0.01,0.0293094)(0.05,0.140116)(0.1,0.265129)(0.15,0.376625)(0.2,0.476089)(0.25,0.564861)(0.3,0.644109)(0.35,0.71479)(0.4,0.777611)(0.45,0.832976)(0.5,0.880923)(0.55,0.921007)(0.6,0.952163)(0.65,0.973012)(0.7,0.98405)(0.75,0.989223)(0.8,0.991862)(0.85,0.993389)(0.9,0.994366)(0.95,0.995037)(1,0.995525)
};
\addplot[mark=diamond,smooth,red,line width=.5pt] plot coordinates {
(0.001,0.00191415)(0.005,0.00952097)(0.01,0.0189185)(0.05,0.0898405)(0.1,0.168674)(0.15,0.237831)(0.2,0.298498)(0.25,0.351737)(0.3,0.398495)(0.35,0.439604)(0.4,0.475795)(0.45,0.507704)(0.5,0.535885)(0.55,0.560825)(0.6,0.582949)(0.65,0.60263)(0.7,0.620193)(0.75,0.635922)(0.8,0.650063)(0.85,0.662831)(0.9,0.67441)(0.95,0.684961)(1,0.694624)
};
\addplot[mark=triangle,smooth,blue,line width=.5pt] plot coordinates {
(0.001,0.00121338)(0.005,0.00603603)(0.01,0.0119957)(0.05,0.0570916)(0.1,0.107698)(0.15,0.152841)(0.2,0.193323)(0.25,0.229785)(0.3,0.262748)(0.35,0.292641)(0.4,0.319829)(0.45,0.344619)(0.5,0.367282)(0.55,0.38805)(0.6,0.407133)(0.65,0.424714)(0.7,0.440957)(0.75,0.456011)(0.8,0.470005)(0.85,0.483057)(0.9,0.495272)(0.95,0.506742)(1,0.517551)
};

\legend{{$\rho=0.99$},{$\rho=0.5$},{$\rho=0.01$}};
\end{axis}
\end{tikzpicture}
}

\caption{\label{fig-4} Convergence of $\Sig(\beta)$ towards $\Sig_{FP}$ (that verifies $\Tr(\Sig_{FP}^{-1})=m$) when $\beta \to 0$ for $N=12$, $m=3$. The criterion used is $C_1(\beta)=\|\Sig(\beta)-\Sig_{FP}\|_F/\|\Sig_{FP}\|_F$.}
\end{center}
\end{figure}